\documentclass[times, 10pt,twocolumn]{article}
\usepackage{latex8}
\usepackage{times}

\usepackage{theorem}
\usepackage{amsmath}
\usepackage{amssymb}

\def\E{\mathcal{E}}

\def\N{\mathcal{N}}

\def\I{\mathcal{I}}
\def\J{\mathcal{J}}
\def\K{\mathcal{K}}
\def\L{\mathcal{L}}

\def\A{{\mathcal{A}}}

\def\X{\mathcal{X}}
\def\Y{\widehat{\X}}
\def\eps{\epsilon}

\def\str{k} 
\def\eps{\epsilon} 
\def\quant{z} 
\def\TV{\text{\scriptsize \em TV}}

\newtheorem{theorem}                          {Theorem}[section]
\newtheorem{lemma}         [theorem] {Lemma}

\newtheorem{claim}         [theorem] {Claim}
\newtheorem{corollary}         [theorem] {Corollary}

{\theorembodyfont{\rmfamily}
\newtheorem{definition} [theorem]{Definition} {\theorembodyfont{\rmfamily}

\newenvironment{proof}{\noindent{ \textbf{Proof:}}} {$\blacksquare$\vskip \belowdisplayskip}

\newenvironment{prevproof}[2]{\noindent {\bf {Proof of {#1}~\ref{#2}:}}}{$\blacksquare$\vskip \belowdisplayskip}

\newcommand{\arxiv}[1]{#1}
\newcommand{\nonarxiv}[1]{}
\newcommand{\shorten}[1]{}

\begin{document}

\title{Discretized Multinomial Distributions\\ and Nash Equilibria in Anonymous Games}

\author{Constantinos Daskalakis\thanks{Supported by a Microsoft Research Fellowship.}~~~~~~~~~~~~~~~~~~
Christos H. Papadimitriou\thanks{The authors were supported through NSF grant CCF - 0635319, a gift from Yahoo! Research, and a
MICRO grant.}\\
University of California, Berkeley\\ Computer Science \\ \{costis, christos\}@cs.berkeley.edu
}

\maketitle
\thispagestyle{empty}

\begin{abstract}
We show that there is a polynomial-time approximation scheme for computing Nash equilibria in
anonymous games with any fixed number of strategies (a very broad
and important class of games), extending the two-strategy result of~\cite{DPanonymous}. The approximation guarantee follows from a
probabilistic result of more general interest: The distribution of
the sum of $n$ independent unit vectors with values ranging over
$\{e_1,\ldots,e_k\}$, where $e_i$ is the unit vector along dimension
$i$ of the $k$-dimensional Euclidean space, can be approximated by
the distribution of the sum of another set of independent unit
vectors whose probabilities of obtaining each value are multiples of~
$1\over z$~~for some integer $z$, and so that the variational
distance of the two distributions is at most~$\epsilon$, where $\epsilon$ is bounded by an inverse
polynomial in $z$ and a function of $k$, but with no dependence on $n$. Our probabilistic result
specifies the construction of a surprisingly sparse $\epsilon$-cover --- under the total variation distance --- of the set of
distributions of sums of independent unit vectors,
which is of interest on its own right.
\end{abstract}

\section{Introduction}

The recent results implying that the Nash equilibrium is an
intractable problem~\cite{DGP}, even in the two-player case~\cite{CD}, have directed the interest of researchers towards
algorithms or complexity results for  special cases~\cite{GaleKuhnTucker,vN,AKV,KT,DP:oblivious}
and approximation algorithms~\cite{LMM,KPS,DMP1,FNS,DMP2,BBM: approximate markakis,ST:approximate 2,DP:oblivious}, and the following
has emerged as the main open question in the area of equilibrium
computation: {\em\bf Is there a PTAS for the Nash
equilibrium?}\footnote{It is shown in \cite{CDT} that an FPTAS is no
more likely than an exact solution.}

In this paper we make progress on this problem, focusing on a very
broad and common class of games called {\em anonymous games}~\cite{Blonskia,Blonskib}. A game is anonymous if the utility of each
player depends not on exactly which other player chooses which
strategy; instead, it only depends on the {\em number} of other
players that play each strategy (that is, it is a symmetric
function of the strategies played by other players). Anonymous games
are a much more general class than the symmetric games (known to be
solvable in polynomial time when the number of strategies is fixed
\cite{PR}), in which all players are identical. Many problems of
interest in computational game theory, such as congestion games,
participation games, voting games, and certain markets and auctions,
are anonymous. Anonymous games have also been used for modeling certain social phenomena~\cite{Blonskib}. Since in anonymous games a player's utility depends
on the {\em partition} of the remaining players into
strategies, such games are a rare case of multiplayer games that
have a polynomially succinct representation --- as long as the
number of strategies is fixed. {\em Our main result is a PTAS
for such games.}  (However, it should be noted that it is not known
whether this special case of the Nash equilibrium problem is PPAD-complete,
and so even an exact algorithm may be possible.)

Our PTAS extends to several generalizations of anonymous games, for
example the case in which there are a few {\em types} of players,
and the utilities depend on how many players {\em of each type} play
each strategy; and to the case in which we have {\em extended
families} (disjoint graphical games of constant degree and with up to
logarithmically many players, each with a utility depending in
arbitrary, possibly non-anonymous, ways on their neighbors, in addition
to their anonymous, possibly typed, interest on everybody else). Essentially any further
extension leads to intractability.

Algorithmic Game Theory aspires to understand the Internet and the
markets it encompasses and creates, and therefore it should focus on
{\em multi-player} games.  We believe that our PTAS is a positive
algorithmic result spanning a vast expanse in this space. However,
because of the tremendous analytical difficulties detailed below,
our algorithm is not practical (as we shall see, the number of
strategies appears, exponentially, in the exponent of the running
time).  It could be, of course, the precursor of more practical
algorithms (in fact, such an algorithm for the two-strategy case
has been recently proposed \cite{D:anonymous3}). But, more importantly, our algorithm
should be seen as compelling computational evidence that there are
very extensive and important classes of common games which are free
of the negative implications of the complexity result in \cite{DGP}.

The basic idea of our algorithm is extremely simple and intuitive
(and in fact it had been noted in the past~\cite{K}): Since we are
looking for mixed strategies (probability distributions, one for
each player, on the set of strategies) that are in equilibrium, we
restrict our search to probability distributions assigning to the
strategies probabilities that are multiples of a fixed fraction,
call it $1\over z$, where $z$ is a large enough natural number.  We
call this process {\em discretization}.  We can then consider each
discrete probability distribution as a separate strategy and look
for (approximate) {\em pure} equilibria in the resulting game (the utilities of
the new game can be computed via dynamic programming). The challenge
is to prove that any mixed Nash equilibrium of the original game has
to be close to some approximate pure Nash equilibrium of the resulting game. For
general games this is not very hard to see (even though it had
apparently escaped the attention of the researchers who first
suggested the discretization method \cite{K}), and this observation
yields a $N^{O\left(\log{\log N\over \epsilon}\right)}$ {\em quasi}-PTAS for
computing Nash equilibria in games in which all players have a fixed
number of strategies, where $N$ is the size of the input (Theorem
\ref{th: quasi-PTAS clique games}; note that this complements the
$N^{O\left({\log N / \epsilon^2}\right)}$ quasi-PTAS of~\cite{LMM} for
games with a fixed number of {\em players\/}).  We also point out
that the discretization method gives the first algorithm for
tree-like graphical games with a fixed number of strategies (for
trees, an initial attempt by~\cite{Kearns: exact} in the
two-strategy case was found to have flaws in~\cite{EGG}, while in
the latter paper a polynomial-time algorithm for graphical games with two
strategies on {\em paths} and {\em cycles} was developed).  Our
algorithm applies to all graphical games with a fixed number of
strategies whose graph is of bounded degree and logarithmically bounded
treewidth.

The discretization method requires polynomial time in the case of
anonymous games, because in this case the search space is no
longer the set of all $n$-tuples of discrete distributions, where $n$ is the
number of players (this is exponential in $n$); instead, via dynamic programming (see the proof of Theorem \ref{thm:mixed strategies}), it can be reduced to the
set of all the ordered partitions of $n$ into $\ell=O((z+1)^{k-1})$ parts,
where $\ell$ is the number of discrete probability distributions
defined above, which is polynomial in $n$, if $k$, the number of strategies, and $z$, the discretization, are fixed.

But proving in this case that the approximation is valid turns out
to be a deep problem. One has to establish a probabilistic lemma
stating that, given a multinomial-sum distribution (the sum of
$k$-dimensional unit vector-valued independent but not necessarily
identically distributed random variables), the probabilities can be rounded to
multiples of $1\over z$ so that the variational distance between the resulting
distribution and the original one depends only on $z$ (and in fact this dependence is
inversely polynomial), and on the dimension $k$ (in an arbitrary
way; the bound we can prove is exponential, and we suspect it is
necessary).  This probabilistic lemma for the case of two strategies
(i.e., for binomial-sum distributions) was proved in \cite{DPanonymous} by
clustering the variables into three classes, depending on how large
their expectation is, and then using results from the probability
literature
\cite{BarbourEtAl:book,BarbourLindvall,Rollin:translatedPoissonApproximations}
to approximate each component binomial-sum distribution (both the
original and the rounded one) by Poisson or shifted Poisson
distributions (depending on the cluster), and finally rounding the
probabilities so that the approximations are close.\vspace{3pt}\\
\indent In the multinomial case, however, no useful approximations are
known;
see, e.g.,~\cite{Barbour} for some obstacles in extending the existent
methods to the multinomial case. Another reason that makes the
binomial case easy is that it is essentially one-dimensional: in the
multinomial case on the other hand, watching the balls in one bin, so to speak, provides
small information about the distribution of the remaining balls in the other bins, because the random vectors are not identically distributed. Our proof is very
involved and indirect, resorting to an alternative sampling of each random vector by
funneling a ball down a probabilistic decision tree with $k-1$ leaves
($k$ is the dimension, or number of strategies), ending up
eventually with a binary choice at the leaves.  This choice can now
be discretized similarly to the binomial case --- albeit with much more effort. The decision tree
topologies become the clusters for the approximation, and their
number (exponential in $k$) appears in the variation distance via a union
bound, and, hence, in the exponent of the running time.  We believe that this probabilistic lemma (Theorem
\ref{theorem:main thm}), and its proof, represent an advance of some
substance in the state of the art in this area of applied
probability.

\arxiv{Our result can be interpreted as constructing a surprisingly sparse cover of the set of multinomial-sum distributions under the total variation distance. Covers of metric spaces have been considered in the literature of approximation algorithms, but we know of no non-trivial result working for the total variation distance or producing a cover of the required sparsity to achieve a polynomial-time approximation scheme for the Nash equilibrium in anonymous games. To show the value of our result in another context, we exhibit a family of non-convex optimization problems arising in economics that can be approximated by means of our probabilistic lemma and for which no efficient algorithm was known before. An application of our result for this family of non-convex optimization problems is a PTAS for finding threat points in repeated anonymous games. These results are discussed in Section~\ref{sec:optimization}.
}

\indent In the balance of this section we provide the necessary definitions. In
the next section we describe the basics of the main result,
including the algorithm and an overview of the proof.  The main
part of the proof of the probabilistic lemma is in Section \ref{sec: main proof}, while
in Section~\ref{sec:extensions} we explore the application of our method to broad generalizations of anonymous games, \nonarxiv{repeated anonymous games,} as well as general (non-anonymous) games and graphical games. \arxiv{In Section~\ref{sec:optimization} we present the application of our result to certain types of non-convex optimization problems.} We conclude with a discussion of problems that remain open.

\vspace{-0.3cm}
\subsection{Definitions and Notation}
\vspace{-0.4cm}
An {\em anonymous game} is a triple
$G=(n,k,\{u^p_i\})$ where $[n]=\{1,\ldots,n\}$, $n\geq 2$, is the set
of players, $[k]=\{1,\ldots,k\}$, $k\geq 2$, is the set of strategies,
and $u^p_i$ with $p\in [n]$ and $i\in[k]$ is the utility of player
$p$ when she plays strategy $i$, a function mapping the set of
partitions $\Pi^k_{n-1}=\{(x_1,\ldots,x_k): x_i \in \mathbb{N}_0
\hbox{{\rm~for all~}} i\in[k], \sum_{i=1}^k x_i = n-1\}$ to the
interval $[0,1]$.~\footnote{In the literature on Nash approximation,
utilities are usually normalized in this way so that the approximation
error is additive.} Our working assumptions are that $n$ is large
and $k$ is fixed; notice that, in this case, anonymous games are
{\em succinctly representable} \cite{PR}, in the sense that their
representation requires specifying $O(n^{k})$ numbers, as opposed
to the $nk^n$ numbers required for general games (arguably, succinct
games are the only multiplayer games that are computationally
meaningful, see \cite{PR} for an extensive discussion of this
point). The convex hull of the set $\Pi^k_{n-1}$ will be denoted by
$\Delta^k_{n-1}= \{(x_1,\ldots,x_k): x_i~\geq~0 \text{ for all } i\in [k],~
\sum_{i=1}^k x_i = n-1\}$.

A {\em pure strategy profile} in such a game is a mapping $S$ from
$[n]$ to $[k]$.  A pure strategy profile $S$ is an {\em
$\epsilon$-approximate pure Nash equilibrium}, where $\epsilon \ge 0$,
if, for all $p\in[n]$, $u^p_{S(p)}(x[S,p])+\epsilon  \geq
u^p_i(x[S,p])$ for all $i\in[k]$, where $x[S,p]\in \Pi^k_{n-1}$ is
the partition $(x_1,\ldots,x_k)$ such that $x_i$ is the number of
players $q\in[n]-\{p\}$ with $S(q)=i$.   


A {\em mixed strategy profile} is a set of $n$ distributions $\{\delta_p\in \Delta^k\}_{p\in[n]}$, where by $\Delta^k$ we denote the $(k-1)$-dimensional simplex, or, equivalently, the set of distributions over $[k]$. A mixed strategy profile is an {\em $\epsilon$-Nash equilibrium} if, for all $p\in[n]$ and $j,j'\in[k]$,
$$E_{\delta_1,\ldots,\delta_n}u^p_j(x) > E_{\delta_1,\ldots,\delta_n}u^p_{j'}(x)+\epsilon \Rightarrow \delta_p(j')=0,$$
where $x$ is drawn from $\Pi^k_{n-1}$ by
drawing $n-1$ random samples from $[k]$ independently according to
the distributions $\delta_q, q\neq p$, and forming the induced
partition.

Similarly, a mixed strategy profile is an {\em $\epsilon$-approximate Nash equilibrium} if, for all $p\in[n]$ and $j\in[k]$,
$E_{\delta_1,\ldots,\delta_n}u^p_i(x)+\epsilon  \geq
E_{\delta_1,\ldots,\delta_n}u^p_j(x)$, where $i$ is drawn from $[k]$
according to $\delta_p$ and $x$ is drawn from $\Pi^k_{n-1}$ as above, by
drawing $n-1$ random samples from $[k]$ independently according to
the distributions $\delta_q, q\neq p$, and forming the induced
partition.

Clearly, an $\epsilon$-Nash equilibrium is also an $\epsilon$-approximate Nash equilibrium, but the converse is not true in general (for an extensive discussion, see~\cite{DGP}). All our positive approximation results are for the stronger notion of the $\epsilon$-Nash equilibrium.

\shorten{\vspace{-0.2cm}}
\section{The Main Result}\label{sec:main results}

The {\em total variation distance} between two distributions
$\mathbb{P}$ and $\mathbb{Q}$ over a finite set $\mathcal{A}$ is
$$||\mathbb{P} - \mathbb{Q}||_{\TV} = \frac{1}{2} \sum_{\alpha \in
\A}{\left|\mathbb{P}(\alpha)-\mathbb{Q}(\alpha)\right|}.$$
Similarly, if $X$ and $Y$ are two random variables ranging over a finite set, their total
variation distance, denoted \nonarxiv{$||X - Y||_{\TV},$}
\arxiv{$$||X - Y||_{\TV},$$}
is defined as the total variation distance between their distributions.
The bulk of the paper is dedicated to proving the following result,
generalizing the one-dimensional ($k=2$) case established in
\cite{DPanonymous}.

\begin{theorem} \label{theorem:main thm}
Let $\{p_{i}\in \Delta^k\}_{i \in [n]}$, and let $\{\X_i \in \mathbb{R}^k\}_{i \in
[n]}$ be a set of independent $\str$-dimensional random unit
vectors such that, for all $i \in [n]$, $\ell \in [\str]$, $\Pr[\X_i = e_{\ell}] = p_{i,\ell}$, where $e_{\ell} \in \mathbb{R}^k$
is the unit vector along dimension $\ell$; also, let $z>0$ be an
integer. Then there exists another set of probability vectors
$\{\widehat{p}_{i} \in
\Delta^{\str}\}_{i \in [n]}$ such that
\begin{enumerate}

\item $|\widehat{p}_{i,\ell} - p_{i,\ell}| = O\left(\frac{1}{\quant}\right)$,
for all $i\in[n], \ell \in [\str]$; \label{cond: closeness}

\item $\widehat{p}_{i,\ell}$ is an integer multiple of
$\frac{1}{2^{\str}}\frac{1}{\quant}$, for all $i\in[n], \ell \in
[\str]$; \label{cond: integrality}

\item if $p_{i,\ell} =0$, then $\widehat{p}_{i,\ell}=0$, for all $i\in[n], \ell \in [\str]$; \label{cond: same support}

\item if $\{\Y_i \in \mathbb{R}^k\}_{i \in [n]}$ is a set of independent random unit vectors such that
$\Pr[\Y_i = e_{\ell}] = \widehat{p}_{i,\ell}$, for all $i\in[n], \ell \in
[\str]$, then
\begin{align}
\left|\left|\sum_{i}{\X_i} - \sum_{i}{\Y_i} \right|\right|_{\TV}=O\left(f(\str){\log{\quant}\over{\quant}^{1/5}}\right) \label{eq: final bound}
\end{align} \label{cond: small variation distance}
and, moreover, for all $j \in [n]$,
\begin{align}
\left|\left|\sum_{i\neq j}{\X_i} - \sum_{i \neq j}{\Y_i} \right|\right|_{\TV}= O\left(f(\str){\log{\quant}\over{\quant}^{1/5}}\right), \label{cond: small variation distance even after removing a guy}
\end{align}
where $f(k)$ is an exponential function of $k$ estimated in the proof.
\end{enumerate}
\end{theorem}

\noindent In other words, there is a way to quantize any set of $n$ independent
random vectors into another set of $n$ independent random vectors, whose
probabilities of obtaining each value are integer multiples of $\eps
\in [0,1]$, so that the total variation distance between the
distribution of the sum of the vectors before and after the
quantization is bounded by $O(f(\str) 2^{\str/6} \eps^{1/6})$. The important, and perhaps surprising, aspect of this bound is the lack of dependence on the number $n$ of random vectors. From this, the main result of this section follows.

\begin{theorem} \label{thm:mixed strategies}
There is a PTAS for the mixed Nash equilibrium problem for anonymous games with a constant number of strategies.
\end{theorem}

\begin{proof}
Consider a mixed Nash equilibrium $(p_1,\ldots, p_n)$.
We claim that the mixed strategy profile $(\widehat{p}_1,\ldots,
\widehat{p}_n)$ specified by Theorem~\ref{theorem:main thm}
constitutes a $O\left(f(\str) z^{-{1\over 6}}\right)$-Nash
equilibrium. Indeed, for every player $i\in [n]$ and
every pure strategy $m \in [k]$ for that player, let us track down the change
in the expected utility of the player for playing strategy $m$ when
the distribution over $\Pi^k_{n-1}$ defined by the $\{p_j\}_{j \neq
i}$ is replaced by the distribution defined by the
$\{\widehat{p}_j\}_{j \neq i}$. It is not hard to see that the
absolute change is bounded by the total variation distance between
the distributions of the random vectors $\sum_{j \neq i}\X_j$ and $\sum_{j \neq
i}\Y_j$, where $\{\X_j\}_{j \neq i}$ are
independent random vectors distributed according to the
distributions $\{p_j\}_{j \neq i}$ and, similarly, $\{\Y_j\}_{j \neq
i}$ are independent random vectors distributed according to the
distributions $\{\widehat{p}_j\}_{j \neq i}$.~\footnote{To establish this bound we use the fact that all utilities lie in $[0,1]$.} Hence, by Theorem
\ref{theorem:main thm}, the change in the utility of the player is
at most $O(f(\str) z^{-{1\over 6}})$, which implies that the
$\widehat{p}_i$'s constitute an $O(f(\str) z^{-{1\over
6}})$-Nash equilibrium of the game.  If we take
$z=\left({f(\str)/ \epsilon}\right)^{6}$, this is a
$\delta$-Nash equilibrium, for $\delta=O(\epsilon)$.

From the previous discussion it follows that there exists a mixed strategy profile $\{\widehat{p}_i\}_i$ which is of the very special kind described by Property~\ref{cond: integrality} in the statement of Theorem~\ref{theorem:main thm} and constitutes a $\delta$-Nash equilibrium of the given game, if we choose $z=\left({f(\str) / \epsilon}\right)^{6}$. The problem is, of course, that we do not know such a mixed strategy profile and, moreover, we cannot afford to do exhaustive search over all mixed strategy profiles satisfying Property~\ref{cond: integrality}, since there is an exponential number of those. We do instead the following search which is guaranteed to find a $\delta$-Nash equilibrium.

Notice that there are at most $(2^kz)^k=2^{k^2}\left({f(\str) / \epsilon}\right)^{6k}$$=:K$ ``quantized'' mixed strategies with each probability being a multiple of $\frac{1}{2^k}\frac{1}{z}$, $z=\left({f(\str)/\epsilon}\right)^{6}$. Let $\K$ be the set of such quantized mixed strategies. We start our algorithm by guessing the partition of the number $n$ of players into quantized mixed strategies; let $\theta=\{\theta_{\sigma}\}_{\sigma \in \K}$ be the partition, where $\theta_{\sigma}$ represents the number of players choosing the discretized mixed strategy $\sigma \in \K$.  Now we only need to determine if there exists an assignment of mixed strategies to the players in $[n]$, with $\theta_{\sigma}$ of them playing mixed strategy $\sigma \in \K$, so that the corresponding mixed strategy profile is a $\delta$-Nash equilibrium. To answer this question it is enough to solve the following {\em max-flow} problem. Let us consider the bipartite graph $([n],\K,E)$ with edge set $E$ defined as follows: $(i,\sigma) \in E$, for $i \in [n]$ and $\sigma \in \K$, if $\theta_{\sigma} >0$ and $\sigma$ is a $\delta$-best response for player $i$, if the partition of the other players into the mixed strategies in $\K$ is the partition $\theta$, with one unit subtracted from $\theta_{\sigma}$.~\footnote{For our discussion, a mixed strategy $\sigma$ of player $i$ is a {\em $\delta$-best response} to a set of mixed strategies for the other players iff the expected payoff of player $i$ for playing any pure strategy $s$ in the support of $\sigma$ is no more than $\delta$ worse than her expected payoff for playing any pure strategy $s'$.} Note that to define $E$ expected payoff computations are required. By straightforward dynamic programming, the expected utility of player $i$ for playing pure strategy $s \in [k]$ given the mixed strategies of the other players can be computed with $O(k n^k)$ operations on numbers with at most $b(n,z,k):=\lceil 1+n (k+\log_2{z}) + \log_2(1/u_{\min}) \rceil$ bits, where $u_{\min}$ is the smallest non-zero payoff value of the game.~\footnote{To compute a bound on the number of bits required for the expected utility computations, note that the expected utility is positive, cannot exceed $1$, and its smallest possible non-zero value is at least $(\frac{1}{2^k}\frac{1}{z})^n u_{\min}$, since the mixed strategies of all players are from the set $\K$. } To conclude the construction of the max-flow instance we add a source node $u$ connected to all the left hand side nodes and a sink node $v$ connected to all the right hand side nodes. We set the capacity of the edge $(\sigma,v)$ equal to $\theta_{\sigma}$, for all $\sigma \in \K$, and the capacity of all other edges equal to $1$. If the max-flow from $u$ to $v$ has value $n$ then there is a way to assign discretized mixed strategies to the players so that $\theta_{\sigma}$ of them play mixed strategy $\sigma \in \K$ and the resulting mixed strategy profile is a $\delta$-Nash equilibrium (details omitted). There are at most $(n+1)^{K-1}$ possible guesses for $\theta$; hence, the search takes overall time
$$O\left((n K k^2n^kb(n,z,k) + p(n+K+2)) \cdot (n+1)^{K-1}\right),$$
where $p(n+K+2)$ is the time needed to find an integral maximum flow in a graph with $n+K+2$ nodes and edge-weights encoded with at most $\lceil \log_2n \rceil$ bits. Hence, the overall time is
$$n^{O\left(2^{k^2}\left({f(\str)
\over\epsilon}\right)^{6k}\right)} \cdot \log_2(1/u_{\min}).$$
\end{proof}

\noindent{\bf Remark:} Theorem \ref{theorem:main thm} can be interpreted as constructing a sparse cover of the set of distributions of sums of independent random unit vectors under the total variation distance. We know of no non-trivial results working for this distance or achieving the same sparsity.

\shorten{\vspace{-0.2cm}}
\subsection{Discussion of Proof Techniques}\label{sec:discussion}

Observe that, from a technical perspective, the $\str>2$ case of Theorem~\ref{theorem:main thm} is inherently different than the $\str = 2$ case, which was shown in~\cite{DPanonymous} (Theorem 3.1). Indeed, when $\str=2$, knowledge of
the number of players who selected their first strategy determines
the whole partition of the number of players into strategies; therefore, in this case the probabilistic experiment is in some sense {\em one-dimensional}. On the other hand, when $\str>2$, knowledge of the
number of ``balls in a bin'', that is the number of players who selected a particular strategy, does not provide full information about
the number of balls in the other bins. This complication would be quite
benign if the vectors $\X_i$ were identically distributed, since in this case the number of balls in a bin would at least characterize precisely the probability
distribution of the number of balls in the other bins (as a multinomial distribution with one bin less and the bin-probabilities appropriately renormalized). But, in our case, the
vectors $\X_i$ are not identically distributed. Hence, already
for $\str=3$ the problem is fundamentally more involved than in the $\str=2$ case.

Indeed, it turns out that obtaining the result for the $\str=2$ case is easier. Here is the intuition: If the expectation of every $\X_i$ at the first bin was small, their sum would be distributed like a Poisson distribution (marginally at that bin); if the expectation of every $\X_i$ was large, the sum would be distributed like a (discretized) Normal distribution.~\footnote{Comparing, in terms of variational distance, a sum of independent Bernoulli random variables to a Poisson or a Normal distribution is an important problem in probability theory. The approximations we use are obtained by applications of {\em Stein's method}~\cite{BarbourChen,BarbourEtAl:book,Rollin:translatedPoissonApproximations}.} So, to establish the result we can do the following (see~\cite{DPanonymous} for details): First, we cluster the $\X_i$'s into those with small and those with large expectation at the first bin, and then we discretize the $\X_i$'s separately in the two clusters in such a way that the sum of their expectations (within each cluster) is preserved to within the discretization accuracy. To show the closeness in total variation distance between the sum of the $\X_i$'s before and after the discretization, we compare instead the Poisson or Normal distributions (depending on the cluster) which approximate the sum of the $\X_i$'s: For the ``small cluster'', we compare the Poisson distributions approximating the sum of the $\X_i$'s before and after the discretization. For the ``large cluster'', we compare the Normals approximating the sum of the $\X_i$'s before and after the discretization.

One would imagine that a similar technique, i.e., approximating by a multidimensional Poisson or Normal distribution, would work for the $\str>2$ case.  Comparing a sum of multinomial random variables to a multidimensional Poisson or Normal distribution is a little harder in many dimensions (see the discussion in \cite{Barbour}), but almost optimal bounds {\em are} known for both the multidimensional Poisson~\cite{Barbour,RoosPoisson} and the multidimensional Normal~\cite{Bha,Goe} approximations. Nevertheless, these results by themselves are not sufficient for our setting: Approximating by a multidimensional Normal performs very poorly at the coordinates where the vectors have small expectations, and approximating by a multidimensional Poisson fails at the coordinates where the vectors have large expectations.  And in our case, it could very well be that the sum of the $\X_i$'s is distributed like a multidimensional Poisson distribution in a subset of the coordinates and like a multidimensional Normal in the complement (those coordinates where the $\X_i$'s have respectively small or large expectations). What we really need, instead, is a multidimensional approximation result that combines the multidimensional Poisson and Normal approximations in the same picture; and such a result is not known. 

%

Our approach instead is very indirect. We define an alternative way
of sampling the vectors $\X_i$ which consists of performing a random
walk on a binary decision tree and performing a probabilistic choice
between two strategies at the leaves of the tree (Sections
\ref{sec:TDP} and \ref{sec:sampling}). The random vectors are then
clustered so that, within a cluster, all vectors share the same
decision tree (Section \ref{sec:clustering}), and the rounding,
performed separately for every cluster, consists of discretizing the
probabilities for the probabilistic experiments at the leaves of the
tree (Section \ref{sec:rounding}). The rounding is done in such a
way that, if all vectors $\X_i$ were to end up at the same leaf
after walking on the decision tree, then the one-dimensional result described above
would apply for the (binary) probabilistic choice that the vectors
are facing at the leaf. However, the random walks will not all end
up at the same leaf with high probability. To remedy this, we define
a coupling between the random walks of the original and the discretized vectors for
which, in the typical case, the probabilistic experiments that the
original vectors will run at every leaf of the tree are very
``similar'' to the experiments that the discretized vectors will run. That is, our coupling guarantees that, with high
probability over the random walks, the total variation distance
between the choices (as random variables) that are to be made by the original
vectors at every leaf of the decision tree and the choices (again as
random variables) that are to be made by the discretized vectors is very small. The
coupling of the random walks is defined in Section~\ref{sec:coupling
in a cell}, and a quantification of the similarity of the leaf
experiments under this coupling is given in Section~\ref{sec:proof of basic lemma}.

For a discussion about why naive approaches such as {\em
rounding to the closest discrete distribution} or {\em randomized rounding} do not appear  useful, even for the $k=2$ case, see Section 3.1 of
\cite{DPanonymous}.

\section{Proof of Theorem \ref{theorem:main thm}} \label{sec: main proof}

\subsection{The Trickle-down Process}\label{sec:TDP}
Consider the mixed strategy $p_i$ of player $i$.  The crux of our
argument is an alternative way to sample from this distribution,
based on the so-called {\em trickle-down process}, defined next.\\

\noindent {\bf TDP} --- Trickle-Down Process\\
\noindent {\bf Input:} $(S,p)$, where $S=\{i_1,\ldots,i_m\}
\subseteq [\str]$ is a set of strategies and $p$ a probability
distribution $p({i_j})>0:j=1,\ldots,m$.  We assume that the elements
of $S$ are ordered $i_1,\ldots,i_m$ in such a way that (a)
$p({i_2})$ is the largest of the $p({i_j})$'s and (b) for $2\neq
j<j'\neq 2$, $p({i_j})\leq p({i_{j'}})$.   That is, the largest
probability is second, and, other than that, the probabilities are
sorted in non-decreasing order (ties broken lexicographically).

\smallskip\noindent ~{\bf if} $|S| \le 2$ stop;\\\noindent~{\bf else} apply the {\em  partition and double operation}:
\shorten{\vspace{-0.15cm}}
\begin{enumerate}

\item let $\ell^* < m$ be the (unique) index such that\\
\text{ }~~~~$\sum_{\ell<\ell^*}{p(i_\ell)} \le \frac{1}{2} \text{ and }\sum_{\ell>\ell^*}{p(i_\ell)} < \frac{1}{2};$

\item Define the sets\\
\text{ }~~~~$S_L = \{i_\ell: \ell \leq \ell^*\}$ and $S_R = \{i_\ell: \ell \geq \ell^*\}$

\item Define the probability distribution $p_L$
such that, for all $\ell < \ell^*$, $p_L(i_\ell)= 2 p(i_\ell)$. Also, let $t:=1- \sum_{\ell=1}^{\ell^*-1}{p_L(i_\ell)}$; if $t=0$, then remove ${\ell^*}$ from $S_L$, otherwise set $p_L(i_{\ell^*})=t$. Similarly, define the probability distribution $p_R$ such that $p_R(i_\ell)= 2 p(i_\ell)$, for all $\ell >
\ell^*$ and $p_R(i_{\ell^*}) = 1- \sum_{\ell^*+1}^m{p_R(i_{\ell})}$.
Notice that, because of the way we have ordered the strategies in
$S$, $i_{\ell^*}$ is neither the first nor the last element of $S$
in our ordering, and hence $2\leq |S_L|, |S_R|<|S|$.

\item call {\bf TDP}$(S_L, p_L)$; call {\bf TDP}$(S_R, p_R)$;
\end{enumerate}

That is, TDP splits the support of the mixed strategy of a player into a tree of finer and
finer sets of strategies, with all leaves having just two
strategies. At each level the two sets in which the set of
strategies is split overlap in at most one strategy (whose probability
mass is divided between its two copies).  The two sets then have
probabilities adding up to~$1/2$, but then the probabilities
are multiplied by $2$, so that each node of the tree represents a
distribution.

\subsection{The Alternative Sampling of $\X_i$} \label{sec:sampling}

Let $p_i$ be the mixed strategy of player $i$, and $\mathcal{S}_i$ be its support.~\footnote{In this section and the following two sections we assume that $|\mathcal{S}_i| >1$; if not, we set $\widehat{p}_i = p_i$, and all claims we make in Sections~\ref{sec:coupling in a cell} and \ref{sec:proof of basic lemma} are trivially satisfied.}  The execution of
{\bf TDP}$(\mathcal{S}_i, p_i)$ defines a rooted binary tree $T_i$ with node
set $V_i$ and set of leaves $\partial T_i$. Each
node $v \in V_i$ is identified with a pair $(S_v, p_{i,v})$, where
$S_v \subseteq [\str]$ is a set of strategies and $p_{i,v}$ is a
distribution over $S_v$.
%
%
%
%
%
Based on this tree, we define the following alternative way to sample $\X_i$:

\medskip\noindent {\bf {\sc Sampling} $\X_i$}
\begin{enumerate}
\item ({\em Stage 1}) Perform a random walk from the root of the tree $T_i$ to the leaves,
where, at every non-leaf node, the left or right child is chosen
with probability $1/2$; let $\Phi_i \in \partial T_i$ be the
(random) leaf chosen by the random walk;

\item ({\em Stage 2}) Let $(S, p)$ be the label assigned to the
leaf $\Phi_i$, where $S =\{\ell_1, \ell_2\}$; set $\X_i =
e_{\ell_1}$, with probability $p({\ell_1})$, and $\X_i =
e_{\ell_2}$, with probability $p({\ell_2})$.
\end{enumerate}

The following lemma, whose straightforward proof we omit, states
that this is indeed an alternative sampling of the mixed strategy of
player $i$.

\begin{lemma}\label{lem: faithful sampling}
For all $i \in [n]$, the process {\sc Sampling}  $\X_i$ outputs
$\X_i = e_\ell$ with probability $p_{i,\ell}$, for all $\ell \in
[\str]$.
\end{lemma}

\subsection{Clustering the Random Vectors}\label{sec:clustering}

We use the process {\bf TDP} to cluster the random vectors of the
set $\{\X_i\}_{i \in [n]}$. We define a cell for every possible tree structure. In particular, for some $\alpha>0$ to be determined later in the proof,

\begin{definition}[Cell Definition]
\noindent {\em Two vectors $\X_i$ and $\X_j$ belong to the same cell
if
\begin{itemize}
\item there exists a tree isomorphism $f_{i,j}: V_i \rightarrow V_j$
between the trees $T_i$ and $T_j$ such that,
for all $u \in V_i$, $v \in V_j$, if $f_{i,j}(u) = v$, then $S_u = S_v$,
and in fact the elements of $S_u$ and $S_v$ are ordered the same way by $p_{i,u}$ and $p_{j,v}$.

\item if $u \in \partial T_i$,
$v=f_{i,j}(u)\in\partial T_j$, and $\ell^* \in S_u=S_v$
is the strategy with the smallest probability mass for both $p_{i,u}$ and
$p_{j,v}$, then either $p_{i,u}(\ell^*), p_{j,v}(\ell^*) \leq {\lfloor z^\alpha\rfloor\over z}$ or $p_{i,u}(\ell^*), p_{j,v}(\ell^*) > {\lfloor z^\alpha\rfloor\over z}$; the leaf is called {\em Type A leaf} in the first case, {\em Type B leaf} in the second case.
\end{itemize}}
\end{definition}
\shorten{\vspace{-0.2cm}}

It is easy to see that the total number of cells is bounded by a function of $\str$ only, estimated in the following claim; \arxiv{the proof of the claim is postponed to Appendix \ref{sec: skipped prooofs}.}\nonarxiv{see the full version of the paper~\cite{DPanonymous2} for the proof of the claim.}

\begin{claim} \label{claim: finite number of cells}
Any tree resulting from TDP has at most $\str-1$ leaves, and the
total number of cells is bounded by $g(k)= \str^{\str^2} 2^{\str-1} 2^{\str} \str!$.
\end{claim}

\subsection{Discretization within a Cell} \label{sec:rounding}

Recall that our goal is to ``discretize'' the probabilities in the
distribution of the $\X_i$'s. We will do this separately in every cell of our clustering. In particular, supposing that $\{\X_i\}_{i \in \I}$ is the set of vectors falling in a particular cell, for some index set $\I$, we will define a set of ``discretized'' vectors
$\{\Y_i\}_{i \in \I}$ in such a way that, for
$h(\str)=\str2^{\str}$, and for all $j\in \I$,
\begin{align}
\left|\left|\sum_{i \in \I}{\X_i} - \sum_{i \in \I}{\Y_i}
\right|\right|_{\TV} =&~O(h(\str)\log{\quant}\cdot{\quant}^{-1/5});
\label{eq:total var within a subcell}\\
\left|\left|\sum_{i \in \I \setminus \{j\}}{\X_i} - \sum_{i \in \I \setminus \{j\}}{\Y_i}
\right|\right|_{\TV} &= O(h(\str)\log{\quant}\cdot{\quant}^{-1/5}). \label{eq:total var within a subcell even after removing a guy}
\end{align}
We establish these bounds in Section \ref{sec:coupling in a cell}. Using the bound on the number of cells in Claim~\ref{claim: finite number of cells}, an
easy application of the coupling lemma implies the bounds shown in~\eqref{eq: final bound} and~\eqref{cond: small variation distance even after removing a guy} for
$f(\str):= h(\str) \cdot g(\str)$, thus concluding the
proof of Theorem \ref{theorem:main thm}.

We shall henceforth concentrate on a particular cell containing the vectors $\{\X_i\}_{i \in {\cal I}}$, for some ${\cal I} \subseteq [n]$. Since the trees $\{T_i\}_{i \in \cal I}$
are isomorphic, for notational convenience we shall
denote all those trees by $T$.  To define the vectors
$\{\Y_i\}_{i \in \I}$ we must provide, for all $i \in \I$, a
distribution $\widehat{p}_i:[\str] \rightarrow [0,1]$ such that
$\Pr[\Y_i = e_{\ell}] = \widehat{p}_i(\ell)$, for all $\ell \in
[\str]$.  To do this, we assign to all $\{\Y_i\}_{i \in {\cal I}}$ the tree $T$ and then, for every leaf $v \in \partial T$ and $i \in \cal I$,
define a distribution $\widehat{p}_{i,v}$ over the two-element
ordered set $S_v$, by the {\sc Rounding} process below.  Then the
distribution $ \widehat{p}_i$ is implicitly defined as
$\widehat{p}_i(\ell) = \sum_{v \in \partial T: \ell \in
S_v}{2^{-depth_T(v)}\widehat{p}_{i,v}(\ell)}.$

\medskip \noindent {\sc Rounding}: for all $v \in \partial T$ with $S_v=\{\ell_1, \ell_2\}$, $\ell_1,\ell_2 \in
[\str]$ do the following
\shorten{\vspace{-0.16 cm}}
\begin{enumerate}
\item find a set of probabilities $\{p_{i,\ell_1}\}_{i \in \I}$ with the following properties \label{rounding: step 2}
\begin{itemize}
\item for all $i\in\I$, $|p_{i,\ell_1} - p_{i,v}(\ell_1)| \le \frac{1}{\quant}$;

\item for all $i\in\I$, $p_{i,\ell_1}$ is an integer multiple of $\frac{1}{\quant}$;

\item $\left|\sum_{i\in \I}{p_{i,\ell_1}} - \sum_{i\in \I}{p_{i,v}(\ell_1)} \right| \le \frac{1}{\quant}$;
\end{itemize}
\item for all $i\in\I$, set $\widehat{p}_{i,v}(\ell_1):=p_{i,\ell_1}$, $\widehat{p}_{i,v}(\ell_2):=1-p_{i,\ell_1}$;
\end{enumerate}

Finding the set of probabilities required by Step \ref{rounding:
step 2} of the {\sc Rounding} process is straightforward and the
details are omitted (see \cite{DPanonymous}, Section 3.3 for a way
to do so). It is now easy to check that the set of probability
vectors $\{\widehat{p}_i\}_{i \in \I}$ satisfies Properties \ref{cond:
closeness}, \ref{cond: integrality} and \ref{cond: same support} of
Theorem \ref{theorem:main thm}.

\subsection{Coupling within a Cell}\label{sec:coupling in a cell}
We are now coming to the main part of the proof:  Showing that the variational distance between the original and the discretized distribution within a cell depends only on $\quant$ and $\str$.  We will only argue that our discretization satisfies~\eqref{eq:total var within a subcell}; the proof of~\eqref{eq:total var within a subcell even after removing a guy} is identical. 

Before proceeding let us introduce some notation. Specifically,
\begin{itemize}
\item let $\Phi_i \in \partial T$ be the leaf chosen by Stage 1 of the process {\sc Sampling} $\X_i$ and $\hat{\Phi}_i \in \partial T$ the leaf  chosen by Stage 1 of {\sc Sampling} $\Y_i$;

\item let $\Phi=\left(\Phi_i\right)_{i\in\I}$ and let $G$ denote the distribution of $\Phi$; similarly, let $\hat{\Phi}=(\hat{\Phi}_i)_{i\in\I}$ and let $\widehat{G}$ denote the distribution of $\hat{\Phi}$.
\end{itemize}

\noindent Moreover, for all $v \in \partial T$, with
${S}_v=\{\ell_1,\ell_2\}$ and ordering $(\ell_1,\ell_2)$,

\begin{itemize}
\item let $\I_{v} \subseteq \I$ be the (random) index set such that $i\in \I_{v}$ iff $i\in\I \wedge \Phi_i = v$ and, similarly, let $\widehat{\I}_{v} \subseteq \I$ be the (random) index set such that $i\in \widehat{\I}_{v}$ iff $i\in\I \wedge \hat{\Phi}_i = v$;

\item let $\J_{v,1}, \J_{v,2} \subseteq \I_{v}$ be the (random) index sets such $i \in \J_{v,1}$ iff $i \in \I_{v} \wedge \X_i =e_{\ell_1}$ and $i \in \J_{v,2}$ iff $i \in \I_{v} \wedge \X_i =e_{\ell_2}$;

\item let $T_{v,1} = |\J_{v,1}|$, $T_{v,2} = |\J_{v,2}|$ and let $F_{v}$ denote the distribution of $T_{v,1}$;

\item let $T:=((T_{v,1}, T_{v,2}))_{v \in \partial T}$ and let $F$ denote the distribution of $T$;

\item let $\widehat{\J}_{v,1}$, $\widehat{\J}_{v,2}$, $\widehat{T}_{v,1}$, $\widehat{T}_{v,2}$, $\widehat{T}$, $\widehat{F}_{v}$, $\widehat{F}$ be defined similarly.
\end{itemize}

The following is easy to see; \nonarxiv{for its proof look at the full version of the paper~\cite{DPanonymous2}.} \arxiv{we postpone its proof to the appendix.}
\begin{claim}\label{claim: phi and psi have same distribution}
For all $\theta \in (\partial T)^{\I}$, $G(\theta) = \widehat{G}(\theta).$
\end{claim}

Since $G$ and $\widehat{G}$ are the same distribution we will
henceforth denote that distribution by $G$. The following lemma is
sufficient to conclude the proof of Theorem \ref{theorem:main thm}.

\begin{lemma} \label{lem: basic lemma}
There exists a value of $\alpha$, used in the definition of the cells, such that, for all $v \in \partial T$,
\begin{align}
&G\left(
\begin{minipage}[h]{7cm}
$\theta:$\\
\text{ }~~$|| F_{v}(\cdot |\Phi = \theta) -  \widehat{F}_{v}(\cdot |\hat{\Phi} = \theta) ||_{\TV} \le O\left(\frac{2^{\str} \log {\quant}}{\quant^{1/5}}\right)$
\end{minipage}
\right) \notag \\&~~~~~~~~~~~~~~~~~~~~~~~~~~~~~~~~~~~~~~~~~~~~~~~~~~~~~~~~~~~~~~~~\ge 1- \frac{4}{\quant^{1/3}}, \label{eq: almost done}
\end{align}
where $F_{v}(\cdot |\Phi)$ denotes the conditional probability distribution of $T_{v,1}$ given $\Phi$ and, similarly, $\widehat{F}_{v}(\cdot |\hat{\Phi} )$ denotes the conditional probability distribution of $\widehat{T}_{v,1}$ given $\hat{\Phi}$.
\end{lemma}

Lemma \ref{lem: basic lemma} states roughly that, for all $v\in
\partial T$, with probability at least $1- \frac{4}{\quant^{1/3}}$ over the
choices made by Stage 1 of processes \{{\sc Sampling} $\X_i$\}$_{i
\in \I}$ and \{{\sc Sampling} $\Y_i$\}$_{i \in \I}$ --- assuming that these processes are coupled to make the same decisions in Stage 1 --- the total
variation distance between the conditional distribution of $T_{v,1}
$ and $\widehat{T}_{v,1}$ is bounded by $O\left(\frac{2^{\str} \log
{\quant}}{\quant^{1/5}}\right)$. The following lemma, whose proof is provided in \nonarxiv{the full version of the paper~\cite{DPanonymous2}}\arxiv{the appendix}, concludes the proof of the main theorem.

\begin{lemma} \label{lemma: done}
\eqref{eq: almost done} implies
\begin{align}||  F- \widehat{F}||_{\TV}  \le O\left(\str\frac{2^{\str} \log {\quant}}{\quant^{1/5}}\right).\label{eq:ha}
\end{align}
\end{lemma}
Note that \eqref{eq:ha} easily implies \eqref{eq:total var within a subcell}

\subsection{Proof of Lemma \ref{lem: basic lemma}} \label{sec:proof of basic lemma}

To conclude the proof of Theorem~\ref{theorem:main thm}, it remains to show Lemma~\ref{lem: basic lemma}. Roughly speaking, the proof consists of showing that, with high probability over the random walks performed in Stage 1 of {\sc Sampling}, the one-dimensional experiment occurring at a particular leaf $v$ of the tree is similar in both the original and the discretized distribution. The similarity is quantified by Lemmas \ref{lemma: concentration for Type A leaves} and \ref{lemma:concentration for leaves of type B} for leaves of type A and B respectively. Then, Lemmas \ref{lemma: if
mus are good everything is good - SMALL}, \ref{lemma: if mus are
good everything is good - BIG and large number} and \ref{lemma: if
mus are good everything is good - BIG and small number} establish
that, if the experiments are sufficiently similar, they can be coupled so that their outcomes agree with high probability.

More precisely, let $v \in \partial T$, $\mathcal{S}_v=\{\ell_1,\ell_2\}$, and suppose the ordering $(\ell_1,\ell_2)$. Also, let us denote
$\ell^*_v = \ell_1$ and define the following functions
\begin{itemize}
\item $\mu_v(\theta) := \sum_{i: \theta_i =v}p_{i,v}(\ell^*_{v})$;

\item $\widehat{\mu}_v(\hat{\theta}) := \sum_{i: \hat{\theta}_i =v }\widehat{p}_{i,v}(\ell^*_{v})$.

\end{itemize}
Note that the random variable $\mu_v(\Phi)$ represents the total probability mass that is placed on the strategy $\ell^*_v$ after the Stage 1 of the {\sc Sampling} process is completed for all vectors $\X_i$, $i \in \I$. Conditioned on the outcome of Stage 1 of {\sc Sampling} for the vectors $\{\X_i\}_{i \in \I}$, $\mu_v(\Phi)$ is the expected number of the vectors from $\I_v$ that will select strategy $\ell^*_v$ in Stage 2 of {\sc Sampling}. Similarly, conditioned on the outcome of Stage 1 of {\sc Sampling} for the vectors $\{\widehat{\X}_i\}_{i \in \I}$, $\widehat{\mu}_v(\widehat{\Phi})$ is the expected number of the vectors from $\widehat{\I}_v$ that will select strategy $\ell^*_v$ in Stage 2 of {\sc Sampling}.

Intuitively, if we can couple the choices made by the random vectors $\X_i$, $i \in \I$, in Stage 1 of {\sc Sampling} with the choices made by the random vectors $\widehat{\X}_i$, $i \in \I$, in Stage 1 of {\sc Sampling} in such a way that, with overwhelming probability, $\mu_v(\Phi)$ and $\widehat{\mu}_v(\widehat{\Phi})$ are close, then also the conditional distributions $F_{v}(\cdot |\Phi)$, $\widehat{F}_{v}(\cdot |\hat{\Phi})$ should be close in total variation distance. The goal of this section is to make this intuition rigorous. We do this in $2$ steps by showing the following.
\begin{enumerate}

\item The choices made in Stage 1 of {\sc Sampling} can be coupled so that the absolute difference $|\mu_v(\Phi) - \widehat{\mu}_v(\widehat{\Phi})|$ is small with high probability. (Lemmas \ref{lemma: concentration for Type A leaves} and \ref{lemma:concentration for leaves of type B}.)\label{step 1}

\item If the absolute difference $|\mu_v(\theta) - \widehat{\mu}_v(\widehat{\theta})|$ is sufficiently small, then so is the total variation distance $|| F_{v}(\cdot |\Phi = \theta) -  \widehat{F}_{v}(\cdot |\hat{\Phi} = \theta) ||_{TV}$. (Lemmas \ref{lemma: if mus are good everything is good - SMALL}, \ref{lemma: if mus are good everything is good - BIG and large number}, and \ref{lemma: if mus are good everything is good - BIG and small number}.) \label{step 2}
\end{enumerate}

We start with Step \ref{step 2} of the above program. We use different arguments depending on whether $v$ is a Type A or Type B leaf. Let $\partial T = \L_A \sqcup \L_B$, where $\L_A$ is the set of type A leaves of the cell and $\L_B$ the set of type B leaves of the cell. For some constant $\beta$ to be decided later, we show the following lemmas.

\begin{lemma} \label{lemma: if mus are good everything is good - SMALL}
For some $\theta \in (\partial T)^{\I}$ and $v \in \L_A$ suppose
that
\begin{align}
\left|\mu_v(\theta) - \E[\mu_v(\Phi)]\right| \le \quant^{(\alpha-1)/2}\sqrt{\E[\mu_v(\Phi)] \log{\quant}} \label{eq: mu X bound}\\
\left|\widehat{\mu}_v(\theta) -
\E[\widehat{\mu}_v(\hat{\Phi})]\right| \le
\quant^{(\alpha-1)/2}\sqrt{\E[\widehat{\mu}_v(\hat{\Phi})]
\log{\quant}} \label{eq: mu Y bound}
\end{align}
then
$$|| F_{v}(\cdot |\Phi = \theta) -  \widehat{F}_{v}(\cdot |\hat{\Phi} = \theta) ||_{TV} \le  O\left(\frac{\sqrt{\log \quant}}{\quant^{(1-\alpha)/2}}\right).$$
\end{lemma}

\begin{lemma} \label{lemma: if mus are good everything is good - BIG and large number}
For some $\theta \in (\partial T)^{\I}$ and $v \in \L_B$ suppose
that
\begin{align}
&n_v(\theta):=|\{i : \theta_i = v \}| \ge \quant^{\beta},\label{condition: }\\
&\left|\mu_v(\theta) - \widehat{\mu}_v(\theta) \right| \le
\frac{1}{\quant} + \frac{\sqrt{\log{\quant}}}{\quant} \sqrt{|\I| },\label{eq: condition type A leaf is good}\\
&|n_v(\theta) - 2^{-depth_T(v)}|\I|| \le \sqrt{3
\log{\quant}}\sqrt{2^{-depth_T(v)}|\I|};\label{eq: number of guys on every leaf concentrated}
\end{align}
then
\begin{align*}&|| F_{v}(\cdot |\Phi = \theta) -  \widehat{F}_{v}(\cdot |\hat{\Phi} = \theta) ||_{TV}\\ &~~~~~~~~~~~\le O\left(\frac{2^{\frac{depth_T(v)}{2}} \sqrt{\log{z}}} {z^{\frac{1+\alpha}{2}}}\right) + O\left(\frac{2^{\frac{depth_T(v)}{2}} {\log{z}}} {z^{\frac{\alpha+\beta+1}{2}}}\right) \\&~~~~~~~~~~~~~~~~~~+O(z^{-\alpha})+O(z^{-({\alpha+\beta -1\over 2})}).
\end{align*}
\end{lemma}

\begin{lemma} \label{lemma: if mus are good everything is good - BIG and small number}
For some $\theta \in (\partial T)^{\I}$ and $v \in \L_B$ suppose
that
\begin{align}
n_v(\theta):=|\{i : \theta_i = v \}| \le \quant^{\beta}
\end{align}
then
$$|| F_{v}(\cdot |\Phi = \theta) -  \widehat{F}_{v}(\cdot |\hat{\Phi} = \theta) ||_{TV} \le O(\quant^{-(1-\beta)}).$$
\end{lemma}

\noindent The proof of Lemma \ref{lemma: if mus are good everything
is good - BIG and small number} follows from a coupling argument
similar to that used in the proof of Lemma 3.13 in~\cite{DPanonymous} and is
omitted. The proofs of Lemmas \ref{lemma: if mus are good everything is good - SMALL} and \ref{lemma: if mus are good everything is good - BIG and large number} \nonarxiv{are given in the full version~\cite{DPanonymous2}.}\arxiv{can be found respectively in Sections \ref{sec:if mus are good everything is good - SMALL} and \ref{sec: if mus are good everything is good - BIG and large number} of the appendix.} Lemma \ref{lemma: if mus are good everything is good - SMALL} provides conditions which, if satisfied by some $\theta$ at a leaf of Type~A, then the conditional distributions $F_{v}(\cdot |\Phi = \theta)$ and $\widehat{F}_{v}(\cdot |\hat{\Phi} = \theta)$ are close in total variation distance. Similarly, Lemmas \ref{lemma: if mus are good everything is good - BIG and large number} and \ref{lemma: if mus are good everything is good - BIG and small number} provide conditions for the leaves of Type B. The following lemmas state that these conditions are satisfied with high probability. Their proof is given in \nonarxiv{the full version~\cite{DPanonymous2}.} \arxiv{Section \ref{sec:concentration of leaf experiments} of the appendix.}

\begin{lemma} \label{lemma: concentration for Type A leaves}
Let $v \in \L_A$. Then
\nonarxiv{
\begin{align}
&G\left(
\begin{minipage}{3.8cm}
$\theta: $ \eqref{eq: mu X bound} and \eqref{eq: mu Y bound} are satisfied\end{minipage}
\right)\ge 1-4 \quant^{-1/3}.
\end{align}
}

\arxiv{
\begin{align}
&G\left(
\begin{minipage}{6.7cm}
$\theta: \left|\mu_v(\theta) - \E[\mu_v(\Phi)]\right| \le
\frac{\sqrt{\log{\quant}}}{\quant^{(1-\alpha)/2}}\sqrt{\E[\mu_v(\Phi)]}$
\\\text{ }~~$\wedge \left|\widehat{\mu}_v(\theta) -
\E[\widehat{\mu}_v(\hat{\Phi})]\right| \le
\frac{\sqrt{\log{\quant}}}{\quant^{(1-\alpha)/2}}\sqrt{\E[\widehat{\mu}_v(\hat{\Phi})]
}$
\end{minipage}
\right)\notag\\&~~~~~~~~~~~~~~~~~~~~~~~~~~~~~~~~~~~~~~~~~~~~~~~~~~~~~~~~~~~\ge 1-4 \quant^{-1/3}.
\end{align}
}
\end{lemma}

\begin{lemma}\label{lemma:concentration for leaves of type B}
Let $v \in \L_B$. Then
\nonarxiv{
\begin{align}
G\left(
\begin{minipage}{4.1cm}
$\theta: $ \eqref{eq: condition type A leaf is good} and \eqref{eq: number of guys on every leaf concentrated} are satisfied
\end{minipage}
\right) \ge 1-
\frac{4}{\quant^{1/2}}.
\end{align}
}

\arxiv{
\begin{align}
&G\left(
\begin{minipage}{7.8cm}
$\theta: \left|\mu_v(\theta) - \widehat{\mu}_v(\theta) \right|
\le \frac{1+\sqrt{|\I|\log{\quant}}}{\quant}~~\wedge$\\$
~~|n_v(\theta) - 2^{-depth_T(v)}|\I|| \le \sqrt{3
\log{\quant}}\sqrt{2^{-depth_T(v)}|\I|}$
\end{minipage}
\right)\notag\\&~~~~~~~~~~~~~~~~~~~~~~~~~~~~~~~~~~~~~~~~~~~~~~~~~~~~~~~~~~~\ge 1-
\frac{4}{\quant^{1/2}}.
\end{align}
}
\end{lemma}

Setting $\alpha=\frac{3}{5}$ and $\beta=\frac{4}{5}$, combining the above, and using that $depth_T(v) \le \str$, as implied by Claim \ref{claim: finite number of cells}, we get~\eqref{eq: almost done}, regardless of whether $v\in\L_A$ or $v \in \L_B$.


\shorten{\vspace{-0.1cm}}\section{Extensions} \label{sec:extensions}
\shorten{\vspace{-0.2cm}}Returning to our algorithm (Theorem \ref{thm:mixed strategies}),
there are several directions in which it can be immediately
generalized. To give an idea of the possibilities, let us define a
{\em semi-anonymous game} to be a game in which
\shorten{\vspace{-0.2cm}}\begin{itemize}
\item the players are partitioned into a fixed number of {\em
types;}

\item there is another partition of the players into an arbitrary number of
disjoint graphical games (see \cite{K}, games in which a node's
utility depends only on its neighboring nodes) of size $O(\log n)$, where $n$ is the total number of players,
and bounded degree called {\em extended families};
\end{itemize}
\shorten{\vspace{-0.2cm}}and the utility of each player depends on (a) his/her own
strategy; (b) the overall number of other players of each type
playing each strategy; and (c) it also depends, in an arbitrary way, on the strategy
choices of neighboring nodes in his/her own extended family.  The
following result, which is only indicative of the applicability of
our approach, can be shown by extending the discretization method
via dynamic programming (details omitted):


\shorten{\vspace{-0.1cm}}
\begin{theorem}\label{thm:semianonymous}
There is a PTAS for semi-anonymous games with a fixed number of
strategies.
\end{theorem}
Further generalizations (for example, not bounding the size of the extended families) lead to PPAD-complete problems.

The discretization approach for the Nash equilibrium that we employed so
far in this paper to anonymous games with a fixed number of
strategies has surprisingly broad applicability, for example
yielding a quasi-PTAS for general games \nonarxiv{(proof in the full version~\cite{DPanonymous2})}\arxiv{(proof in Appendix \ref{sec: skipped prooofs})}:

\begin{theorem} \label{th: quasi-PTAS clique games}
In any normal-form game with a constant number of strategies per player, an $\epsilon$-approximate Nash equilibrium can be
computed in time $N^{O(\log{\log N\over \epsilon})}$, where $N$ is
the description size of the game.
\end{theorem}

By combining the discretization approach with the techniques of
\cite{DP:pure nash via markov} we can find approximate Nash
equilibria in a large class of {\em graphical games}.  It had long
been thought that graphical games on trees with two strategies per
player can be solved in polynomial time \cite{Kearns: exact}, until
subtle flaws in the algorithm were discovered \cite{EGG}.  The
largest class of graphical games that are known to have a polynomial-time
algorithm for Nash equilibria is graphical games on a cycle and two
strategies per player \cite{EGG}.  The following result treats a far
broader class of games, albeit approximately; its proof is omitted.
\shorten{\vspace{-0.2cm}}
\begin{theorem}
There is PTAS for computing Nash equilibria in graphical games in
which each player has a number of strategies bounded by a constant
and the graph has bounded degree and $O(\log n)$ treewidth.
\end{theorem}

\nonarxiv{
Our probabilistic lemma has also been used for solving repeated anonymous games via the so-called ``Folk Theorem''. This result for repeated games follows as a special case of a family of non-convex optimization problems for which our probabilistic lemma provides a PTAS (see the discussion in the full version of the paper~\cite{DPanonymous2}).
\begin{theorem}[\cite{folk}]
There is a PTAS for computing threat points in repeated anonymous games with a constant number of strategies per player.
\end{theorem}

}

\arxiv{
\section{An Application to Optimization}\label{sec:optimization}

We illustrate an interesting application of our method in non-convex optimization. This application relates nicely to the interpretation of our main result (Theorem \ref{theorem:main thm}) as constructing a sparse cover of the set of distributions of sums of independent unit vectors under the total variation distance. The minimax optimization problem that we present arises in the context of solving repeated anonymous games, using the folk theorem \cite{folk}, and similar optimization problems arise naturally in economics whenever secure strategies or threats are being computed. The optimization problem that we consider is the following.

\medskip
\begin{minipage}[h]{7.5cm}
\noindent {\em
Given functions $f_1, f_2 : \{0,1,\ldots,n\} \rightarrow [0,1]$ solve the optimization problem
\begin{align}\min_{p_1,\ldots, p_n \in [0,1]} \max_{k \in \{1,2\}} \left\{ \E_{X_i \sim B(p_i)} \left[ f_k\left(\sum_{i=1}^n X_i\right) \right] \right\}, \label{eq: maximin}
\end{align}
where $\E_{X_i \sim B(p_i)}$ denotes the expectation over the joint measure of independent Bernoulli random variables $X_i, i=1,\ldots,n$, with expectations $p_i, i=1,\ldots,n$.
}
\end{minipage}

\medskip We know of no efficient algorithm for solving the above optimization problem. Nevertheless, our technique gives rise to a polynomial time approximation scheme. The idea is to use Theorem \ref{theorem:main thm} to show that restricting the search space from $[0,1]^n$ to $\{0,\epsilon, 2\epsilon,\ldots, 1\}^n$ results in a loss of at most $O(\epsilon^{1/6})$ in the value of the optimum. This observation is complemented by the symmetry of the objective function with respect to the $p_i$'s; hence, we can search over the discretized space in time $O(n^{1/\epsilon})$ rather than $(1/\epsilon)^n$. The proof of the following theorem is given in detail in Appendix \ref{sec: skipped prooofs}.

\begin{theorem} \label{thm: minimax}
There is a PTAS for solving the non-convex optimization problem \eqref{eq: maximin}.
\end{theorem}

The algorithm extends to the case that the minimax problem is replaced by a maximin problem. Moreover, our method provides polynomial time approximation schemes for several generalizations of \eqref{eq: maximin}, e.g., for the case that the maximum is taken over more than two functions (the case of one function is trivial), the domain of the functions is multidimensional (but with a constant number of dimensions), the functions have several (but constant number of) arguments, etc. Theorem \ref{thm: minimax} implies immediately the following result.

\begin{corollary}[\cite{folk}]
There is a PTAS for computing threat points in repeated anonymous games with a constant number of strategies per player.
\end{corollary}
}

\section{Open Problems}\label{sec:open problems}

\shorten{\vspace{-0.2cm}}Is there a PTAS for the Nash equilibrium problem? A major progress
in this direction would be to turn the quasi-PTAS we described in
the previous section for the case of a fixed number of strategies to
a true PTAS.  This is challenging, of course, but not hopeless.  The
exhaustive algorithm need not be completely exhaustive; a more
intelligent search of the space, possibly in a dynamically varying
grid of discretized probabilities,  could possibly bring
improvements in the running time.  On the other hand, any constant
lower bound on the approximability would be great progress as well;
we conjecture that such a bound is possible at least for graphical
games.

Obviously, our PTAS is not ready to be implemented and run; the
exponent makes it unrealistic for any reasonable $\epsilon$.  (As we have argued in the
Introduction, its true significance lies in delimiting the
implications of the complexity result in \cite{DGP}.)  There are
ways to improve it, perhaps even substantially.  For example, by a
more elaborate trickle-down process all trees could be made full
binary trees of depth $\log k$, which would remove one of the exponential functions from the exponent of the
running time.  But a truly practical algorithm
would have to start from a new idea --- possibly from that of a
``less exhaustive search'' mentioned in the previous paragraph. In fact, an efficient PTAS for the case of two strategies has been recently suggested~\cite{D:anonymous3}.

\arxiv{
\newpage
\onecolumn
\begin{appendix}

\section*{APPENDIX}
\section{Skipped Proofs} \label{sec: skipped prooofs}

\begin{prevproof}{Claim}{claim: finite number of cells}
That a tree resulting from TDP has $k-1$ leaves follows by
induction: It is true when $k=2$, and for general $k$, the left
subtree has $j$ strategies and thus, by induction, $j-1$ leaves, and
the right subtree has at most $k+1-j$ strategies and $k-j$ leaves; adding we
get the result.

To estimate the number of cells, let us fix the set of strategies and their ordering at the root of the tree (thus the result of the calculation will have to be multiplied by $2^kk!$) and then count the number of trees that could be output by TDP. Suppose that the root has cardinality $m$ and that the children of the root are assigned sets of sizes $j$ and $m+1-j$ (or, in the event of no duplication, $m-j$), respectively. If $j=2$, then a duplication has to have happened and, for the ordering of the strategies at the left child of the root, there are at most $2$ possibilities depending on whether the ``divided strategy'' is still the largest at the left side; similarly, for the right side there are $m-1$ possibilities: either the divided strategy is still the largest at the right side, or it is not in which case it has to be inserted at the correct place in the ordering and the last strategy of the right side must be moved to the second place. If $j > 2$, similar considerations show that there are at most $j-1$ possibilities for the left side and $1$ possibility for the right side. It follows that the number of trees is
bounded from above by the solution $T(k)$ of the recurrence
\begin{align*}
&T(n)=2~T(2)\cdot (n-1)T(n-1)\\&~~~~~+\sum_{j=3}^{n-1}(j-1)T(j)\cdot \max\{T(n-j),T(n+1-j)\}.
\end{align*}
with T(2)=1.  It follows that the total number of trees can be
upper-bounded by the function $k^{k^2}$. Taking into account that there are $2^kk!$ choices for the set of strategies and their ordering at the root of the tree, and that each leaf can be of either Type A, or of Type B, it follows that the total number of cells is bounded by $g(k)= k^{k^2} 2^{k-1} 2^k k!$.
\end{prevproof}

\begin{prevproof}{Claim}{claim: phi and psi have same distribution}
The proof follows by a straightforward coupling argument. Indeed,  for all $i \in \I$,
let us couple the choices made by Stage 1 of {\sc Sampling} $\X_i$ and {\sc Sampling}
$\Y_i$ so that the random leaf $\Phi_i \in \partial T$ chosen by
{\sc Sampling} $\X_i$ and the random leaf $\hat{\Phi}_i \in \partial T$ chosen by
{\sc Sampling} $\Y_i$ are equal, that is, for all $i\in \I$, in the joint probability space
$\Pr[\Phi_i = \hat{\Phi}_i]=1$; the existence of such a coupling is straightforward
since Stage 1 of both {\sc Sampling} $\X_i$ and {\sc Sampling}
$\Y_i$ is the same random walk on $T$.
\end{prevproof}

\begin{prevproof}{Lemma}{lemma: done}
Note first that \eqref{eq: almost done} implies via a union bound that
\begin{align}
G\left(
\theta : \forall v \in \partial T, || F_{v}(\cdot |\Phi =
\theta) -  \widehat{F}_{v}(\cdot |\hat{\Phi} = \theta) ||_{\TV}\le O\left(\frac{2^{\str} \log {\quant}}{\quant^{1/5}}\right)
\right) \ge
1- O(\str \quant^{-1/3}), \label{eq: all the leaves are close}
\end{align}
since, by Claim~\ref{claim: finite number of cells}, the number of leaves is at most $k-1$. 

Now suppose that, for a given value of $\theta \in (\partial T)^{\I}$, the following is satisfied
\begin{align}
\forall v \in \partial T,~|| F_{v}(\cdot |\Phi = \theta) -
\widehat{F}_{v}(\cdot |\hat{\Phi} = \theta) ||_{\TV} \le
O\left(\frac{2^{\str} \log {\quant}}{\quant^{1/5}}\right).
\label{eq:all the leaves are close 2}
\end{align}
Observe that the variables $\{T_{v,1} \}_{v\in \partial T}$ are conditionally independent given $\Phi$, and, similarly, the variables $\{\widehat{T}_{v,1} \}_{v\in \partial T}$ are conditionally independent given $\hat{\Phi}$.
This, by the coupling lemma, Claim \ref{claim: finite number of cells}, and~\eqref{eq:all the leaves are close 2} implies that
\begin{align*}
|| F(\cdot |\Phi = \theta) -  \widehat{F}(\cdot |\hat{\Phi} = \theta) ||_{\TV} \le O\left(\str\frac{2^{\str} \log {\quant}}{\quant^{1/5}}\right),
\end{align*}
where we used that, if $\Phi = \hat{\Phi}$, then $|\I_v|=|\widehat{\I}_v|$, $\forall v \in\partial T$.

Therefore, \eqref{eq: all the leaves are close} implies
\begin{align}
&G\left(
\theta: ||F(\cdot |\Phi = \theta) -  \widehat{F}(\cdot |\hat{\Phi} = \theta) ||_{\TV} \le O\left(\str\frac{2^{\str} \log {\quant}}{\quant^{1/5}}\right)
\right) \ge 1- O(\str \quant^{-1/3}). \label{eq:
important}
\end{align}

All that remains is to shift the bound of \eqref{eq:
important} to the unconditional space. The following lemma
establishes this reduction.

\begin{lemma} \label{claim: total var from conditional}
\eqref{eq: important} implies
\begin{align}||  F- \widehat{F}||_{\TV}  \le O\left(\str\frac{2^{\str} \log {\quant}}{\quant^{1/5}}\right).
\end{align}
\end{lemma}

\begin{prevproof}{Lemma}{claim: total var from conditional} Let us denote by
$$Good = \{\theta| \theta\in(\partial T)^{\I} : ||F(\cdot |\Phi = \theta) -  \widehat{F}(\cdot |\hat{\Phi} = \theta) ||_{\TV} \le O\left(\str\frac{2^{\str} \log {\quant}}{\quant^{1/5}}\right),$$
and let $Bad = (\partial T)^{\I} - Good$. By~\eqref{eq: important}, if follows that $G(Bad) \le O(\str \quant^{-1/3})$.

\begin{align*}
||T - \widehat{T}||_{\TV}  &= \frac{1}{2} \sum_{ t} | F(t) - \widehat{F}(t)|\\
&= \frac{1}{2} \sum_{ t } \left| \sum_{\theta} F(t|\Phi = \theta) G(\Phi = \theta) - \sum_{\theta} \widehat{F}(t|\hat{\Phi}=\theta) \widehat{G}(\hat{\Phi}=\theta) \right|\\
&= \frac{1}{2} \sum_{t} \left| \sum_{\theta} (F(t|\Phi=\theta)  -  \widehat{F}({t}|\hat{\Phi}=\theta)) G(\theta) \right|~~~~~~~~~\left(\text{using $G(\theta)=\widehat{G}(\theta),\forall \theta$}\right)\\
&\le \frac{1}{2} \sum_{ t } \sum_{\theta}\left| F(t|\Phi=\theta)  -  \widehat{F}(t|\hat{\Phi}=\theta) \right|G(\theta)\\
&= \frac{1}{2} \sum_{ t} \sum_{\theta \in Good}\left| F(t|\Phi=\theta)  -  \widehat{F}(t|\hat{\Phi}=\theta) \right|G(\theta)\\
&~~~~~~~~~+\frac{1}{2} \sum_{ t } \sum_{\theta \in Bad}\left| F(t|\Phi=\theta)  -  \widehat{F}(t|\hat{\Phi}=\theta) \right|G(\theta)\\
&\le \sum_{\theta \in Good} G(\theta) \left(\frac{1}{2} \sum_{ t} \left| F(t|\Phi=\theta)  -  \widehat{F}(t|\hat{\Phi}=\theta) \right|\right)\\
&~~~~~~~~~+\sum_{\theta \in Bad}G(\theta)\left(\frac{1}{2} \sum_{ t } \left| F(t|\Phi=\theta)  -  \widehat{F}(t|\hat{\Phi}=\theta) \right|\right)\\
&\le \sum_{\theta \in Good} G(\theta) \cdot O\left(\str\frac{2^{\str} \log {\quant}}{\quant^{1/5}}\right) +\sum_{\theta \in Bad}G(\theta)\\
&\le O\left(\str\frac{2^{\str} \log {\quant}}{\quant^{1/5}}\right) + O(\str \quant^{-1/3}).
\end{align*}

\end{prevproof}\end{prevproof}

\begin{prevproof}{Theorem}{th: quasi-PTAS clique games}
Let $p$ be the number of players and $s$ the number of strategies per player which we assume to be a constant; the input size is $N=ps^p$.  Consider a new $p$-player game in which the set of pure strategies of each player is the set of all distributions over the $s$ strategies of the original game whose probabilities are integer multiples of $\delta={\epsilon\over 2ps}$. We claim that, if we search over all the pure strategy profiles of the new game, we are bound to discover an $\epsilon$-approximate Nash equilibrium of the original game. To prove this, it suffices to notice the following which is proven below by two applications of the coupling lemma.

\begin{lemma} \label{lem:petrubation lemma}
Let $(x_1,\ldots,x_p)$ be a mixed-Nash equilibrium of the original game and $(\hat{x}_1,\ldots,\hat{x}_p)$ be another set of mixed strategies, where, for all $i$, $j$, $\hat{x}_i(j)$ is an integer multiple of $\delta={\epsilon \over 2ps}$, $|\hat{x}_i(j)-{x}_i(j)| \le \delta$ and, if $x_i(j)=0$, then also $\hat{x}_i(j)$=0. Then $(\hat{x}_1,\ldots,\hat{x}_p)$ is an $\epsilon$-approximate Nash equilibrium of the original game.
\end{lemma}
The number of pure strategy profiles of the new game that we have to search over is at most  $\left(\left( {1 \over \delta} \right)^{s}\right) ^p$, which is easily seen to be $N^{O(\log{\log N\over \epsilon})}$.
\end{prevproof}

\begin{prevproof}{Lemma}{lem:petrubation lemma}
For every player $i$ and strategy $j$, let $U^i_j$ and $\hat{U}^i_j$ be the expected utility of player $i$ if she plays $j$ and the other players play $\{x_{i'}\}_{i' \neq i}$ and $\{\hat{x}_{i'}\}_{i' \neq i}$ respectively. The difference between $U^i_j$ and $\hat{U}^i_j$ can be bounded as follows
$$|{U}^i_j-\hat{U}^i_j| \le ||(x_1,\ldots,x_p) - (\hat{x}_1,\ldots,\hat{x}_p)||_{\TV},$$
where the right hand side of the above expression represents the total variation distance between the compound distributions $(x_1,\ldots,x_p)$ and $(\hat{x}_1,\ldots,\hat{x}_p)$, and we used the fact that the payoff functions of the players lie in $[0,1]$. We will show that
$$||(x_1,\ldots,x_p) - (\hat{x}_1,\ldots,\hat{x}_p)||_{\TV} \le {\epsilon \over 2}.$$
Indeed, for all $i$, let $\X_i$ be a random $s$-dimensional vector such that $\X_i = e_j$ with probability $x_i(j)$, and suppose that the vectors $\{\X_i\}_{i}$ are independent. Similarly, define vectors $\{\hat{\X}_i\}_i$. The coupling lemma implies that, for any coupling of $\{\X_i\}_{i}$ and $\{\hat{\X}_i\}_{i}$,
$$||(\X_1,\ldots, \X_p) - (\hat{\X}_1,\ldots,\hat{\X}_p)||_{\TV} \le \Pr[(\X_1,\ldots, \X_p) \neq (\hat{\X}_1,\ldots, \hat{\X}_p)],$$
which, by a union bound, implies
$$||(\X_1,\ldots, \X_p) - (\hat{\X}_1,\ldots,\hat{\X}_p)||_{\TV} \le \sum_i \Pr[\X_i \neq \hat{\X}_i].$$
Let us now fix a coupling for which, for all $i$,
$$\Pr[\X_i \neq \hat{\X}_i] = ||\X_i - \hat{\X}_i||_{\TV}.$$
Such a coupling exists by the coupling lemma and the fact that the random vectors $\{\X_i\}_{i}$ are independent and so are the random vectors $\{\hat{\X}_i\}_{i}$. Combining the above, we get
$$||(\X_1,\ldots, \X_p) - (\hat{\X}_1,\ldots,\hat{\X}_p)||_{\TV} \le \sum_i ||\X_i - \hat{\X}_i||_{\TV}.$$
Observe finally that
$$||(x_1,\ldots,x_p) - (\hat{x}_1,\ldots,\hat{x}_p)||_{\TV} = ||(\X_1,\ldots, \X_p) - (\hat{\X}_1,\ldots,\hat{\X}_p)||_{\TV},$$
and, for all $i$,
$$\text{} ||\X_i - \hat{\X}_i||_{\TV} = ||x_i - \hat{x}_i||_{\TV} \le \delta s = {\epsilon \over 2p}.$$
It follows that $$||(x_1,\ldots,x_p) - (\hat{x}_1,\ldots,\hat{x}_p)||_{\TV} \le {\epsilon \over 2}.$$
Hence, for all $i$, $j$,
$$|{U}^i_j-\hat{U}^i_j| \le ||(x_1,\ldots,x_p) - (\hat{x}_1,\ldots,\hat{x}_p)||_{\TV} \le {\epsilon \over 2},$$
which implies that $(\hat{x}_1,\ldots,\hat{x}_p)$ is an $\epsilon$-approximate Nash equilibrium of the original game.
\end{prevproof}

\begin{prevproof}{Theorem}{thm: minimax}
It is not hard to see that for any sets of probabilities $\{p_i\}_{i}$ and $\{p_i'\}_{i}$, and for any $\alpha \in \{1,2\}$,

$$\left| \E_{X_i \sim B(p_i)} \left[ f_{\alpha}\left(\sum_{i=1}^n X_i\right) \right] - \E_{Y_i \sim B(p_i')} \left[ f_{\alpha}\left(\sum_{i=1}^n Y_i\right) \right]\right| \le \left| \left| \sum_{i}{X}_i - \sum_{i}{Y}_i \right| \right|_{\TV},$$
where, in the right hand side of the above $\{X_i\}_i$ is a set of independent Bernoulli random variables with expectations $\{p_i\}_i$ and $\{Y_i\}_i$ a set of independent Bernoulli random variables with expectations $\{p_i'\}_i$.

Suppose now that $\{p^*_i\}_{i}$ is the set of probabilities achieving the optimum value for \eqref{eq: maximin}. It follows from the above that if we perturb the $p^*_i$'s to another set of probabilities $\{p'^*_i\}_{i}$ the value of the minmax problem is only affected by an additive term $\left| \left| \sum_{i}{{X}_i} - \sum_{i}{{Y}_i} \right| \right|_{\TV}$, where ${X}_i \sim B(p^*_i)$ and ${Y}_i \sim B(p'^*_i)$, for all $i$.

It follows from Theorem \ref{theorem:main thm} that, for any set of probabilities $\{p^*_i\}_i$, there exists another set of $\epsilon$-``discretized'' probabilities $\{p'^*_i \}_i$, that is, $p'^*_i$ is an integer multiple of $\epsilon$, for all $i$, such that

$$\left| \left| \sum_{i}{{X}_i} - \sum_{ i}{{Y}_i} \right| \right|_{\TV} \le O(\epsilon^{1/6}).$$

Hence, we can restrict the optimization to $\epsilon$-discretized probabilities with an additive loss of $O( \epsilon^{1/6})$ in the value of the optimum. Even so, the search space is of size $\Omega\left(\left(\frac{1}{\epsilon}\right)^{n}\right)$ which is exponential in the input size $O(n)$. By observing that the objective function is symmetric with respect to the set of probabilities $\{p_i\}_i$ we can prune the search space to searching only over the partitions of $n$ unlabeled objects into $1/\epsilon$ bins, that is $O(n^{1/\epsilon})$ possible partitions. This results in a polynomial time approximation scheme.
\end{prevproof}

\section{Proof of Lemma \ref{lemma: if mus are good everything is good - SMALL}} \label{sec:if mus are good everything is good - SMALL}

\begin{proof}
By the assumption it follows that

\begin{align*}\left|\mu_v(\theta) - \widehat{\mu}_v(\theta) \right| \le \left|\E[\mu_v(\Phi)] - \E[\widehat{\mu}_v(\hat{\Phi})]\right| +\quant^{(\alpha-1)/2}\sqrt{\E[\mu_v(\Phi)] \log{\quant}} + \quant^{(\alpha-1)/2}\sqrt{\E[\widehat{\mu}_v(\hat{\Phi})] \log{\quant}}.
\end{align*}

Moreover, note that
$$\E[\mu_v(\Phi)] = 2^{-depth_T(v)} \sum_{i \in \I}{p_{i,v}(\ell^*_v)}$$
and, similarly,
$$\E[\widehat{\mu}_v(\hat{\Phi})] = 2^{-depth_T(v)} \sum_{i \in \I}{\widehat{p}_{i,v}(\ell^*_v)}.$$
By the definition of the {\sc Rounding} procedure it follows that
$$|\E[\mu_v(\Phi)] - \E[\widehat{\mu}_v(\hat{\Phi})]| \le 2^{-depth_T(v)} \frac{1}{\quant}.$$
Hence it follows that
\begin{align}
\left|\mu_v(\theta) - \widehat{\mu}_v(\theta) \right| \le
2^{-depth_T(v)} \frac{1}{\quant}
+\frac{2\sqrt{\log{\quant}}}{\quant^{(1-\alpha)/2}}\sqrt{\max{\{\E[\mu_v(\Phi),\E[\widehat{\mu}_v(\hat{\Phi})]\}}]
}.\label{eq: mu X - mu Y bound}
\end{align}
\text{}\\

Let $\N_v(\theta):=\{i : \theta_i = v \}$, $n_v = |\N_v|$.
Conditioned on $\Phi = \theta$, the distribution of $T_{v,1}$ is the
sum of $n_v$ independent Bernoulli random variables $\{Z_i\}_{i \in
\N_v}$ with expectations $\E[Z_i] = p_{i,v}(\ell^*_v) \le
\frac{\lfloor \quant^{\alpha}\rfloor}{\quant}$. Similarly,
conditioned on $\hat{\Phi} = \theta$, the distribution of
$\widehat{T}_{v,1}$ is the sum of $n_v$ independent Bernoulli random
variables $\{\widehat{Z}_i\}_{i \in \N_v}$ with expectations
$\E[\widehat{Z}_i] = \widehat{p}_{i,v}(\ell^*_v) \le \frac{\lfloor
\quant^{\alpha}\rfloor}{\quant}$. Note that
$$\E\left[\sum_{i \in \N_v}Z_i\right] = \mu_v(\theta)$$
and, similarly,
$$\E\left[\sum_{i \in \N_v}\widehat{Z}_i\right] = \widehat{\mu}_v(\theta).$$

\noindent Without loss of generality, let us assume that $\E[\mu_v(\Phi)] \ge \E[\widehat{\mu}_v(\hat{\Phi})]$. Let us further distinguish two cases for some constant $\tau < 1-\alpha$ to be decided later\\

\noindent {\bf Case 1: $\E[\mu_v(\Phi)] \le \frac{1}{\quant^{\tau}}$}.\\

From \eqref{eq: mu X bound} it follows that,
$$\mu_v(\theta) \le \E[\mu_v(\Phi)] + \quant^{(\alpha-1)/2}\sqrt{\E[\mu_v(\Phi)] \log{\quant}} \le \frac{1}{\quant^{\tau}} + \frac{\sqrt{\log \quant}}{\quant^{(\tau+1-\alpha)/2}}=:g(\quant).$$
Similarly, because $\E[\widehat{\mu}_v(\hat{\Phi})] \le
\E[\mu_v(\Phi)] \le \frac{1}{\quant^{\tau}}$,
$\widehat{\mu}_v(\theta) \le g(\quant)$.

By Markov's inequality, $\Pr_{\Phi=\theta} [\sum_{i \in \N_v}Z_i \ge
1] \le \frac{\mu_v(\theta)}{1}\le g(\quant)$ and, similarly,
$\Pr_{\hat{\Phi}=\theta} [\sum_{i \in \N_v}\widehat{Z}_i \ge 1] \le
g(\quant)$. Hence,
\begin{align*}
\left|\Pr\text{}_{\Phi=\theta} \left[\sum_{i \in \N_v}Z_i =0\right] - \Pr\text{}_{\hat{\Phi}=\theta} \left[\sum_{i \in \N_v}\widehat{Z}_i = 0\right]\right|&=\left|\Pr\text{}_{\Phi=\theta} \left[\sum_{i \in \N_v}Z_i \ge 1\right] - \Pr\text{}_{\hat{\Phi}=\theta} \left[\sum_{i \in \N_v}\widehat{Z}_i \ge 1\right]\right|\\
&\le 2g(\quant).
\end{align*}
It follows then easily that
\begin{align}
|| F_{v}(\cdot |\Phi = \theta) -  \widehat{F}_{v}(\cdot |\hat{\Phi}
= \theta) ||_{\TV} \le 4 g(\quant) = 4\cdot\left(
\frac{1}{\quant^{\tau}} + \frac{\sqrt{\log
\quant}}{\quant^{(\tau+1-\alpha)/2}}\right).\label{total variation
if expectation super small}
\end{align}

\noindent {\bf Case 2: $\E[\mu_v(\Phi)] \ge \frac{1}{\quant^{\tau}}$}.\\

The following claim was proven in \cite{DPanonymous}, Lemma 3.9,
\begin{claim}
For any set of independent Bernoulli random variables $\{Z_i\}_i$
with expectations $\E[Z_i] \le \frac{\lfloor
\quant^{\alpha}\rfloor}{\quant}$,
$$\left\| \sum_i Z_i - Poisson\left(\E\left(\sum_i Z_i\right)\right) \right\|_{\TV} \le \frac{1}{\quant^{1-\alpha}}.$$
\end{claim}
By application of this lemma it follows that
\begin{align}
\left\| \sum_{i \in \N_v} Z_i - Poisson( \mu_v(\theta) ) \right\|_{\TV} \le \frac{1}{\quant^{1-\alpha}}, \label{eq: sum Z x versus Poisson}\\
\left\| \sum_{i \in \N_v} \widehat{Z}_i - Poisson(
\widehat{\mu}_v(\theta) ) \right\|_{\TV} \le
\frac{1}{\quant^{1-\alpha}}. \label{eq: sum Z y versus Poisson}
\end{align}
We study next the distance between the two Poisson distributions. We
use the following lemma whose proof is postponed till later in this section.
\begin{lemma}\label{lemma:distance between poissons}
If $\lambda=\lambda_0+D$ for some $D>0$, $\lambda_0>0$,
$$\| Poisson(\lambda)  -  Poisson(\lambda_0) \|_{\TV} \le D\sqrt{\frac{2}{\lambda_0}}.$$
\end{lemma}

An application of Lemma \ref{lemma:distance between poissons} gives
\begin{align}
\| Poisson(\mu_v(\theta) )  -  Poisson(\widehat{\mu}_v(\theta) )
\|_{\TV} \le |\mu_v(\theta) - \widehat{\mu}_v(\theta)
|\sqrt{\frac{2}{\min{\{\mu_v(\theta), \widehat{\mu}_v(\theta)\}}}}.
\label{eq: variation bound for our poissons}
\end{align}
We conclude with the following lemma proved in the end of this section.
\begin{lemma}\label{lemma: variation bound for our poissons}
From \eqref{eq: mu X bound}, \eqref{eq: mu Y bound}, \eqref{eq: mu X
- mu Y bound} and  the assumption $\E[\mu_v(\Phi)] \ge
\frac{1}{\quant^{\tau}}$, it follows that
$$|\mu_v(\theta) - \widehat{\mu}_v(\theta) |\sqrt{\frac{2}{\min{\{\mu_v(\theta), \widehat{\mu}_v(\theta)\}}}} \le \sqrt{72\frac{\log{\quant}}{\quant^{1-\alpha}}}.$$
\end{lemma}

Combining \eqref{eq: sum Z x versus Poisson}, \eqref{eq: sum Z y
versus Poisson}, \eqref{eq: variation bound for our poissons} and
Lemma \ref{lemma: variation bound for our poissons} we get
$$\left\| \sum_{i \in \N_v} Z_i - \sum_{i \in \N_v} \widehat{Z}_i \right\|_{\TV} \le \frac{2}{\quant^{1-\alpha}}+\sqrt{72\frac{\log{\quant}}{\quant^{1-\alpha}}} = O\left(\frac{\sqrt{\log \quant}}{\quant^{(1-\alpha)/2}}\right),$$
which implies
\begin{align}|| F_{v}(\cdot |\Phi = \theta) -  \widehat{F}_{v}(\cdot |\hat{\Phi} = \theta) ||_{\TV} \le O\left(\frac{\sqrt{\log \quant}}{\quant^{(1-\alpha)/2}}\right).\label{total variation if expectation small but not super small}
\end{align}
Taking $\tau > (1-\alpha)/2$, we get from \eqref{total variation if
expectation super small}, \eqref{total variation if expectation
small but not super small} that in both cases
\begin{align}|| F_{v}(\cdot |\Phi = \theta) -  \widehat{F}_{v}(\cdot |\hat{\Phi} = \theta) ||_{\TV} \le O\left(\frac{\sqrt{\log \quant}}{\quant^{(1-\alpha)/2}}\right).
\end{align}
\end{proof}

\begin{prevproof}{lemma}{lemma:distance between poissons}
We make use of the following lemmas.
\begin{lemma}
If $\lambda,\lambda_0>0$, the Kullback-Leibler divergence between $Poisson(\lambda_0)$ and $Poisson(\lambda)$ is given by
$$\Delta_{KL}(Poisson(\lambda) || Poisson(\lambda_0)) = \lambda \left( 1 - \frac{\lambda_0}{\lambda} + \frac{\lambda_0}{\lambda} \log\frac{\lambda_0}{\lambda}\right).$$
\end{lemma}
\begin{lemma}[e.g. \cite{CoverThomas}]\label{lemma: relation between total var and KL}
If $P$ and $Q$ are probability measures on the same measure space and $P$ is absolutely continuous with respect to $Q$ then
$$\| P - Q\|_{\TV} \le \sqrt{2 \Delta_{KL}(P||Q)}.$$
\end{lemma}

\noindent By simple calculus we have that
\begin{align*}
\Delta_{KL}(Poisson(\lambda) || Poisson(\lambda_0)) &= \lambda \left( 1 - \frac{\lambda_0}{\lambda} + \frac{\lambda_0}{\lambda} \log\frac{\lambda_0}{\lambda}\right) \le \frac{D^2}{\lambda_0}.
\end{align*}
Then by Lemma \ref{lemma: relation between total var and KL} it follows that
$$\| Poisson(\lambda)  -  Poisson(\lambda_0) \|_{\TV} \le D\sqrt{\frac{2}{\lambda_0}}.$$
\end{prevproof}

\begin{prevproof}{lemma}{lemma: variation bound for our poissons}
From \eqref{eq: mu X - mu Y bound} and the assumption $\E[\mu_v(\Phi)] \ge \E[\widehat{\mu}_v(\hat{\Phi})]$ we have
\begin{align*}
|\mu_v(\theta) - \widehat{\mu}_v(\theta) |^2 &\le \frac{1}{\quant^2} + \frac{4 \log{\quant}}{\quant^{1-\alpha}}\E[\mu_v(\Phi)] + 4 \frac{1}{\quant} \frac{\sqrt{\log{\quant}}}{\quant^{(1-\alpha)/2}}\sqrt{\E[\mu_v(\Phi)]}.
\end{align*}
From the assumption $\E[\mu_v(\Phi)] \ge \frac{1}{\quant^{\tau}}$ it follows
\begin{align}
\E[\mu_v(\Phi)] &= \sqrt{\E[\mu_v(\Phi)]} \sqrt{\E[\mu_v(\Phi)]}\\
&\ge \frac{1}{\quant^{\tau/2}} \sqrt{\E[\mu_v(\Phi)]}.
\end{align}
Since $\tau < 1-\alpha$, it follows that, for sufficiently large $\quant$ which only depends on $\alpha$ and $\tau$, $\frac{1}{\quant^{\tau/2}} \ge \frac{2\sqrt{\log\quant}}{\quant^{(1-\alpha)/2}}$. Hence,
$$\E[\mu_v(\Phi)] \ge \frac{2\sqrt{\log\quant}}{\quant^{(1-\alpha)/2}} \sqrt{\E[\mu_v(\Phi)] },$$
which together with \eqref{eq: mu X bound} implies
\begin{align}
\mu_v(\theta) &\ge  \E[\mu_v(\Phi)] - \quant^{(\alpha-1)/2}\sqrt{\E[\mu_v(\Phi)] \log{\quant}}\ge\frac{1}{2}\E[\mu_v(\Phi)]
 \label{eq: mu X lower bound}
\end{align}
Similarly, starting from $\E[\widehat{\mu}_v(\hat{\Phi})] \ge \E[\mu_v(\Phi)] -\frac{1}{\quant} \ge \frac{1}{\quant^{\tau}} - \frac{1}{\quant}$, it can be shown that for sufficiently large $\quant$
\begin{align}
\widehat{\mu}_v(\theta) &\ge \frac{1}{2}\E[\widehat{\mu}_v(\hat{\Phi})].
 \label{eq: mu Y lower bound}
\end{align}
From \eqref{eq: mu X lower bound}, \eqref{eq: mu Y lower bound} it follows that
\begin{align*}
\min\{\mu_v(\theta),\widehat{\mu}_v(\theta)\} &\ge \frac{1}{2}\min\{\E[\mu_v(\Phi)],\E[\widehat{\mu}_v(\hat{\Phi})]\}=\frac{1}{2}\E[\widehat{\mu}_v(\Phi)] \ge \frac{1}{2}\E[\mu_v(\Phi)]-\frac{1}{2\quant} \ge \frac{1}{4}\E[\mu_v(\Phi)],
\end{align*}
where we used that $\E[\mu_v(\Phi)] \ge \frac{1}{\quant^{\tau}} \ge \frac{2}{\quant}$ for sufficiently large $\quant$, since $\tau < 1-\alpha$. Combining the above we get
\begin{align*}
\frac{2|\mu_v(\theta) - \widehat{\mu}_v(\theta) |^2}{\min{\{\mu_v(\theta), \widehat{\mu}_v(\theta)\}}}
&\le2 \frac{\frac{1}{\quant^2} + \frac{4 \log{\quant}}{\quant^{1-\alpha}}\E[\mu_v(\Phi)] + 4 \frac{1}{\quant} \frac{\sqrt{\log{\quant}}}{\quant^{(1-\alpha)/2}}\sqrt{\E[\mu_v(\Phi)]}}{\frac{1}{4}\E[\mu_v(\Phi)]}\\
&\le 8 \frac{1}{\quant^2 \E[\mu_v(\Phi)]} + 32\frac{\log{\quant}}{\quant^{1-\alpha}} + 32 \frac{\sqrt{\log{\quant}}}{\quant^{1+(1-\alpha)/2}\sqrt{\E[\mu_v(\Phi)]}}\\
&\le 8 \frac{\quant^{\tau}}{\quant^2} + 32\frac{\log{\quant}}{\quant^{1-\alpha}} + 32 \frac{\quant^{\tau/2}\sqrt{\log{\quant}}}{\quant^{1+(1-\alpha)/2}}\\
&\le 8 \frac{1}{\quant^{2-\tau}} + 32\frac{\log{\quant}}{\quant^{1-\alpha}} + 32 \frac{\sqrt{\log{\quant}}}{\quant^{(3-\alpha-\tau)/2}}\\
&\le 72\frac{\log{\quant}}{\quant^{1-\alpha}},
\end{align*}
since $2-\tau > 1-\alpha$ and $(3-\alpha-\tau)/2>1-\alpha$, assuming sufficiently large $z$.
\end{prevproof}

\section{Proof of Lemma \ref{lemma: if mus are good everything is good - BIG and large number}} \label{sec: if mus are good everything is good - BIG and large number}

\begin{proof} We will derive our bound by approximating with the {\em translated Poisson distribution}, which is defined next.

\begin{definition}[\cite{Rollin:translatedPoissonApproximations}] We say that an integer random variable $Y$ has a {\em translated Poisson distribution} with paremeters $\mu$ and $\sigma^2$ and write
$$\L(Y)=TP(\mu,\sigma^2)$$
if $\L(Y - \lfloor \mu-\sigma^2\rfloor) = Poisson(\sigma^2+
\{\mu-\sigma^2\})$, where $\{\mu-\sigma^2\}$ represents the
fractional part of $\mu-\sigma^2$.
\end{definition}\text{}

\noindent The following lemma provides a bound for the total
variation distance between two translated Poisson distributions with
different parameters.

\begin{lemma}[\cite{BarbourLindvall}] \label{lem: variation distance between translated Poisson distributions}
Let  $\mu_1, \mu_2 \in \mathbb{R}$ and $\sigma_1^2, \sigma_2^2 \in
\mathbb{R}_+ \setminus \{0\}$ be such that $\lfloor \mu_1-\sigma_1^2
\rfloor \le \lfloor \mu_2-\sigma_2^2 \rfloor$. Then
\begin{align*}&\left|\left|TP(\mu_1,\sigma_1^2)-TP(\mu_2,\sigma_2^2)\right|\right|_{\TV} \le \frac{|\mu_1-\mu_2|}{\sigma_1}+\frac{|\sigma_1^2-\sigma_2^2|+1}{\sigma_1^2}.\end{align*}
\end{lemma}

The following lemma was proven in \cite{DPanonymous}, Lemma 3.14,
\begin{lemma}\label{lem: translated poisson approximation}
Let $\quant>0$ be some integer and $\{Z_i\}_{i=1}^m$, where $m \ge
\quant^{\beta}$, be any set of independent Bernoulli random
variables with expectations $\E[Z_i] \in \left[\frac{\lfloor
\quant^{\alpha}\rfloor}{\quant},\frac{1}{2}\right]$. Let $\mu_1=
\sum_{i=1}^m\E[Z_i]$ and $\sigma_1^2 =
\sum_{i=1}^m\E[Z_i](1-\E[Z_i])$. Then
$$\left\| \sum_{i=1}^m Z_i - TP\left(\mu_1,\sigma_1^2\right) \right\|_{\TV} \le O\left( {\quant^{-\frac{\alpha+\beta-1}{2}}}\right).$$
\end{lemma}

Let $\N_v(\theta):=\{i : \theta_i = v \}$, $n_v(\theta) =
|\N_v(\theta)|$. Conditioned on $\Phi = \theta$, the distribution of
$T_{v,1}$ is the sum of $n_v(\theta)$ independent Bernoulli random
variables $\{Z_i\}_{i \in \N_v(\theta)}$ with expectations $\E[Z_i]
= p_{i,v}(\ell^*_v)$. Similarly, conditioned on $\hat{\Phi} =
\theta$, the distribution of $\widehat{T}_{v,1}$ is the sum of
$n_v(\theta)$ independent Bernoulli random variables
$\{\widehat{Z}_i\}_{i \in \N_v(\theta)}$ with expectations
$\E[\widehat{Z}_i] = \widehat{p}_{i,v}(\ell^*_v)$. Note that
$$\sum_{i \in \N_v(\theta)}\E\left[Z_i\right] = \mu_v(\theta)$$
and, similarly,
$$\sum_{i \in \N_v(\theta)}\E\left[\widehat{Z}_i\right] = \widehat{\mu}_v(\theta).$$

\noindent Setting $\mu_1 :=  \mu_v(\theta)$, $\mu_2 :=
\widehat{\mu}_v(\theta)$ and
$$\sigma_1^2 = \sum_{i \in \N_v(\theta)}\E\left[Z_i\right](1-\E\left[Z_i\right]),$$
$$\sigma_2^2 = \sum_{i \in \N_v(\theta)}\E\left[\widehat{Z}_i\right](1-\E\left[\widehat{Z}_i\right]),$$
we have from Lemma \ref{lem: translated poisson approximation} that
\begin{align}
\left\| \sum_{i \in \N_v(\theta)} Z_i -
TP\left(\mu_1,\sigma_1^2\right) \right\|_{\TV} \le O\left(
{\quant^{-\frac{\alpha+\beta-1}{2}}}\right). \label{eq: first sum
versus translated poisson}
\end{align}
\begin{align}
\left\| \sum_{i \in \N_v(\theta)} \widehat{Z}_i -
TP\left(\mu_2,\sigma_2^2\right) \right\|_{\TV} \le O\left(
{\quant^{-\frac{\alpha+\beta-1}{2}}}\right). \label{eq: second sum
versus translated poisson}
\end{align}
It remains to bound the total variation distance between the
translated poisson distributions using Lemma \ref{lem: variation
distance between translated Poisson distributions}. Without loss of
generality let us assume $\lfloor \mu_1-\sigma_1^2 \rfloor \le
\lfloor \mu_2-\sigma_2^2 \rfloor$. Note that
$$\sigma_1^2 = \sum_{i \in \N_v(\theta)}\E\left[Z_i\right](1-\E\left[Z_i\right]) \ge n_v(\theta) \frac{\lfloor \quant^{\alpha}\rfloor}{\quant} \left(1- \frac{\lfloor \quant^{\alpha}\rfloor}{\quant}\right) \ge \frac{1}{2}n_v(\theta) \frac{\lfloor \quant^{\alpha}\rfloor}{\quant},$$
where the last inequality holds for values of $z$ which are larger
than some function of constant $\alpha$. Also,
\begin{align*}
|\sigma_1^2-\sigma_2^2| &\le \sum_{i \in \N_v(\theta)} \left| \E\left[Z_i\right] (1-\E\left[Z_i\right]) - \E\left[\widehat{Z}_i\right](1-\E\left[\widehat{Z}_i\right])\right|\\
&= \sum_{i \in \N_v(\theta)} \left| p_{i,v}(\ell^*_v) (1-p_{i,v}(\ell^*_v)) - \widehat{p}_{i,v}(\ell^*_v)(1-\widehat{p}_{i,v}(\ell^*_v))\right|\\
&= \sum_{i \in \N_v(\theta)} (\left| p_{i,v}(\ell^*_v) - \widehat{p}_{i,v}(\ell^*_v) \right| + \left| p_{i,v}^2(\ell^*_v) - \widehat{p}^2_{i,v}(\ell^*_v) \right|)\\
&\le \sum_{i \in \N_v(\theta)} \frac{3}{\quant}~~~~~~~~~~~~~~~~~~~\left(\text{using $\left| p_{i,v}(\ell^*_v) - \widehat{p}_{i,v}(\ell^*_v) \right|\le \frac{1}{\quant}$}\right)\\
&\le \frac{3n_v(\theta)}{\quant}.
\end{align*}

Using the above and Lemma \ref{lem: variation distance between
translated Poisson distributions} we have that
\begin{align*}
\left|\left|TP(\mu_1,\sigma_1^2)-TP(\mu_2,\sigma_2^2)\right|\right| & \le \frac{|\mu_1-\mu_2|}{\sigma_1}+\frac{|\sigma_1^2-\sigma_2^2|}{\sigma_1^2} + \frac{1}{\sigma_1^2}\\
& \le \frac{|\mu_1-\mu_2|}{\sigma_1}+\frac{\frac{3n_v(\theta)}{\quant}}{\frac{1}{2}n_v(\theta) \frac{\lfloor \quant^{\alpha}\rfloor}{\quant}} + \frac{1}{\frac{1}{2}n_v(\theta) \frac{\lfloor \quant^{\alpha}\rfloor}{\quant}}\\
& \le \frac{|\mu_1-\mu_2|}{\sigma_1}+O(z^{-\alpha}) + \frac{1}{\frac{1}{2}\quant^{\beta} \frac{\lfloor \quant^{\alpha}\rfloor}{\quant}}\\
& \le \frac{|\mu_1-\mu_2|}{\sigma_1}+O(z^{-\alpha}) +
O(z^{-(\alpha+\beta -1)}).
\end{align*}
To bound the ratio $\frac{|\mu_1-\mu_2|}{\sigma_1}$ we distinguish
the following cases:
\begin{itemize}
\item $\sqrt{3\log{\quant}}\sqrt{2^{-depth_T(v)}} \sqrt{|\I|} \le \frac{1}{2} 2^{-depth_T(v)} |\I|$: Combining this inequality with \eqref{eq: number of guys on every leaf concentrated}
we get that
$$|\I| \le 2^{1+depth_T(v)} n_v(\theta).$$

Hence,
$$\frac{|\mu_1-\mu_2|}{\sigma_1} \le \frac{\frac{1}{\quant} + \frac{\sqrt{\log{\quant}}}{\quant} \sqrt{|\I| }}{\sqrt{\frac{1}{2}n_v(\theta) \frac{\lfloor \quant^{\alpha}\rfloor}{\quant}}} \le \frac{\frac{1}{\quant} + \frac{\sqrt{\log{\quant}}}{\quant} \sqrt{2^{1+depth_T(v)} n_v(\theta) }}{\sqrt{\frac{1}{2}n_v(\theta) \frac{\lfloor \quant^{\alpha}\rfloor}{\quant}}} = O\left(\frac{1}{z^{\frac{\alpha+\beta+1}{2}}}\right) + O\left(\frac{2^{\frac{depth_T(v)}{2}} \sqrt{\log{z}}} {z^{\frac{1+\alpha}{2}}}\right)$$

\item $\sqrt{3\log{\quant}}\sqrt{2^{-depth_T(v)}} \sqrt{|\I|} > \frac{1}{2} 2^{-depth_T(v)} |\I|$: It follows that
$$|\I| < 12~ 2^{depth_T(v)} \log{z}.$$
Hence,
$$\frac{|\mu_1-\mu_2|}{\sigma_1} \le \frac{\frac{1}{\quant} + \frac{\sqrt{\log{\quant}}}{\quant} \sqrt{|\I| }}{\sqrt{\frac{1}{2}n_v(\theta) \frac{\lfloor \quant^{\alpha}\rfloor}{\quant}}} \le \frac{\frac{1}{\quant} + \frac{\sqrt{\log{\quant}}}{\quant} \sqrt{12~ 2^{depth_T(v)} \log{z}}}{\sqrt{\frac{1}{2}n_v(\theta) \frac{\lfloor \quant^{\alpha}\rfloor}{\quant}}} = O\left(\frac{1}{z^{\frac{\alpha+\beta+1}{2}}}\right) + O\left(\frac{2^{\frac{depth_T(v)}{2}} {\log{z}}} {z^{\frac{\alpha+\beta+1}{2}}}\right)$$
\end{itemize}
Combining the above, it follows that
\begin{align*}
\left|\left|TP(\mu_1,\sigma_1^2)-TP(\mu_2,\sigma_2^2)\right|\right| &\le \frac{|\mu_1-\mu_2|}{\sigma_1}+\frac{|\sigma_1^2-\sigma_2^2|+1}{\sigma_1^2}\\
&\le O\left(\frac{1}{z^{\frac{\alpha+\beta+1}{2}}}\right) + O\left(\frac{2^{\frac{depth_T(v)}{2}} \sqrt{\log{z}}} {z^{\frac{1+\alpha}{2}}}\right) + O\left(\frac{2^{\frac{depth_T(v)}{2}} {\log{z}}} {z^{\frac{\alpha+\beta+1}{2}}}\right)+O(z^{-\alpha}) + O(z^{-(\alpha+\beta -1)})\\
&\le O\left(\frac{2^{\frac{depth_T(v)}{2}} \sqrt{\log{z}}}
{z^{\frac{1+\alpha}{2}}}\right) +
O\left(\frac{2^{\frac{depth_T(v)}{2}} {\log{z}}}
{z^{\frac{\alpha+\beta+1}{2}}}\right)+O(z^{-\alpha})+O(z^{-(\alpha+\beta -1)}).
\end{align*}
Combining the above with \eqref{eq: first sum versus translated
poisson} and \eqref{eq: second sum versus translated poisson} we get
$$\left\| \sum_{i \in \N_v(\theta)} Z_i -   \sum_{i \in \N_v(\theta)} \widehat{Z}_i \right\|_{\TV} \le O\left(\frac{2^{\frac{depth_T(v)}{2}} \sqrt{\log{z}}} {z^{\frac{1+\alpha}{2}}}\right) + O\left(\frac{2^{\frac{depth_T(v)}{2}} {\log{z}}} {z^{\frac{\alpha+\beta+1}{2}}}\right)+O(z^{-\alpha})+O(z^{-({\alpha+\beta -1\over 2})}).$$
\end{proof}

\section{Concentration of the Leaf Experiments}\label{sec:concentration of leaf experiments}
The following lemmas constitute the last piece of the puzzle and
complete the proof of Lemma \ref{lem: basic lemma}. They roughly
state that, after the random walk in Stage 1 of the processes {\sc
Sampling} is performed, the experiments that will take place in
Stage 2 of the processes {\sc Sampling} are similar with high
probability.

\begin{prevproof}{Lemma}{lemma: concentration for Type A leaves}
Note that
$$\mu_v(\Phi) = \sum_{i \in \I} \Omega_i=:\Omega,$$
where $\{\Omega_i\}_i$ are independent random variables defined as
\begin{align*}
\Omega_i =
\begin{cases}
p_{i,v}(\ell^*_v),~~~~\text{with probability }2^{-depth_T(v)} \\
0,~~~~~~~~~~~~~~~\text{with probability }1-2^{-depth_T(v)}.
\end{cases}
\end{align*}
We apply the following version of Chernoff/Hoeffding bounds to the
random variables $\Omega^{'}_i := \quant^{1-\alpha}\Omega_i \in
[0,1]$.
\begin{lemma}[Chernoff/Hoeffding] \label{lemna: chernoff}
Let $Z_1,\ldots, Z_m$ be independent random variables with $Z_i \in
[0,1]$, for all $i$. Then, if $Z=\sum_{i=1}^nZ_i$ and $\gamma \in
(0,1)$,
$$\Pr[|Z-\E[Z]| \ge \gamma \E[Z]] \le 2 \exp(-\gamma^2\E[Z]/3).$$
\end{lemma}
Letting $\Omega^{'}=\sum_{i\in I}\Omega^{'}_i$ and applying the
above lemma with $\gamma:= \sqrt{\frac{1}{\E[\Omega^{'}]}
\log{\quant}}$, it follows that
$$\Pr\left[\left|\Omega^{'} - \E[\Omega^{'}]\right| \ge \sqrt{\E[\Omega^{'}] \log{\quant}} \right] \le 2 \quant^{-1/3},$$
which in turn implies
$$\Pr\left[\left|\Omega - \E[\Omega]\right| \ge \quant^{(\alpha-1)/2}\sqrt{\E[\Omega] \log{\quant}} \right] \le 2 \quant^{-1/3},$$
or, equivalently,
$$\Pr\left[\left|\mu_v(\Phi) - \E[\mu_v(\Phi)]\right| \ge \quant^{(\alpha-1)/2}\sqrt{\E[\mu_v(\Phi)] \log{\quant}} \right] \le 2 \quant^{-1/3}.$$
Similarly, it can be derived that
$$\Pr\left[\left|\widehat{\mu}_v(\hat{\Phi}) - \E[\widehat{\mu}_v(\hat{\Phi})]\right| \ge \quant^{(\alpha-1)/2}\sqrt{\E[\widehat{\mu}_v(\hat{\Phi})] \log{\quant}} \right] \le 2 \quant^{-1/3}.$$
Let us consider the joint probability space which makes $\Phi =
\hat{\Phi}$ with probability $1$; this space exists since as we
observed above $G(\theta)=\widehat{G}(\theta), \forall \theta$. By a
union bound for this space
$$\Pr\left[\left|\mu_v(\Phi) - \E[\mu_v(\Phi)]\right| \ge \frac{\sqrt{\log{\quant}}}{\quant^{(1-\alpha)/2}}\sqrt{\E[\mu_v(\Phi)]} \vee \left|\widehat{\mu}_v(\hat{\Phi}) - \E[\widehat{\mu}_v(\hat{\Phi})]\right| \ge \frac{\sqrt{\log{\quant}}}{\quant^{(1-\alpha)/2}}\sqrt{\E[\widehat{\mu}_v(\hat{\Phi})] }\right] \le 4 \quant^{-1/3}.$$
which implies
$$G\left(\theta: \left|\mu_v(\theta) - \E[\mu_v(\Phi)]\right| \le \frac{\sqrt{\log{\quant}}}{\quant^{(1-\alpha)/2}}\sqrt{\E[\mu_v(\Phi)]} \wedge \left|\widehat{\mu}_v(\theta) - \E[\widehat{\mu}_v(\hat{\Phi})]\right| \le \frac{\sqrt{\log{\quant}}}{\quant^{(1-\alpha)/2}}\sqrt{\E[\widehat{\mu}_v(\hat{\Phi})] } \right) \ge 1-4 \quant^{-1/3}.$$
\end{prevproof}

\begin{prevproof}{Lemma}{lemma:concentration for leaves of type B}
Suppose that the random variables $\Phi$ and $\hat{\Phi}$ are
coupled so that, with probability $1$, $\Phi = \hat{\Phi}$. Then

$$\mu_v(\Phi) - \widehat{\mu}_v(\hat{\Phi}) = \sum_{i \in \I} \Omega_i=:\Omega,$$
where $\{\Omega_i\}_i$ are independent random variables defined as
\begin{align*}
\Omega_i =
\begin{cases}
p_{i,v}(\ell^*_v)-\widehat{p}_{i,v}(\ell^*_v),~~~~\text{with probability }2^{-depth_T(v)} \\
0,~~~~~~~~~~~~~~~\text{with probability }1-2^{-depth_T(v)}.
\end{cases}
\end{align*}

\noindent We apply Hoeffding's inequality to the random variables
$\Omega_i$.

\begin{lemma}[Hoeffding's Inequality]
Let $X_1,\ldots,X_n$ be independent random variables. Assume that,
for all $i$, $\Pr[X_i \in [a_i, b_i]] = 1$. Then, for $t>0$:
 $$\Pr\left[\sum_iX_i - \E\left[\sum_iX_i\right] \geq t\right]\leq \exp \left( - \frac{2 t^2}{\sum_{i=1}^n (b_i - a_i)^2} \right).$$
\end{lemma}
Applying the above lemma we get

$$\Pr \left[  \left| \Omega  - \E\left[\Omega\right]  \right| \ge t \right] \le 2 \exp\left(-\frac{2t^2}{ |\I| \frac{4}{\quant^2}}\right),$$
since, for all $i\in\I$,
$|p_{i,v}(\ell^*_v)-\widehat{p}_{i,v}(\ell^*_v)| \le
\frac{1}{\quant}$. Setting $t= \sqrt{\log{\quant}} \sqrt{|\I|}
\frac{1}{\quant} $ we get
$$\Pr \left[  \left| \Omega  - \E\left[\Omega\right]  \right| \ge \sqrt{\log{\quant}} \sqrt{|\I|} \frac{1}{\quant} \right] \le 2 \frac{1}{\quant^{1/2}}.$$
Note that $$|\E[\Omega]|=|\sum_{i \in \I}{\E[\Omega_i]}| =
|{2^{-depth_T(v)}\sum_{i \in
\I}(p_{i,v}(\ell^*_v)-\widehat{p}_{i,v}(\ell^*_v)})| \le
\frac{1}{\quant}.$$ It follows from the above that
$$\Pr\left[|\Omega| \le \frac{1}{\quant} + \sqrt{\log{\quant}} \sqrt{|\I|} \frac{1}{\quant}\right] \ge 1-2\frac{1}{\quant^{1/2}},$$
which gives immediately that
$$G\left(\theta: \left|\mu_v(\theta) - \widehat{\mu}_v(\theta) \right| \le \frac{1}{\quant} + \frac{\sqrt{\log{\quant}}}{\quant} \sqrt{|\I| }\right) \ge 1- \frac{2}{\quant^{1/2}}.$$
Moreover, an easy application of Lemma \ref{lemna: chernoff} gives
\begin{align} \label{eq: concentration of the number of guys on every leaf}
G\left( \theta: |n_v(\theta) - 2^{-depth_T(v)}|\I|| \le \sqrt{3
\log{\quant}}\sqrt{2^{-depth_T(v)}|\I|} \right) \ge 1- \frac{2}{z}.
\end{align}
Indeed, let $T_i = 1_{\Phi_i = v}$. Then $n_v(\Phi)=\sum_{i \in
\I}T_i$ and $\E[\sum_{i \in \I}T_i] = 2^{-depth_T(v)} |\I|$.
Applying Lemma \ref{lemna: chernoff} with $\gamma =
\sqrt{\frac{3\log z }{2^{-depth_T(v)} |\I|}}$ we get
$$\Pr\left[\left|\sum_{i \in \I}T_i - \E\left[\sum_{i \in \I}T_i\right] \right| \ge \sqrt{3 \log{\quant}}\sqrt{2^{-depth_T(v)}|\I|}\right] \le \frac{2}{z},$$
which implies
$$\Pr\left[|n_v(\Phi) - 2^{-depth_T(v)}|\I|| \le \sqrt{3 \log{\quant}}\sqrt{2^{-depth_T(v)}|\I|} \right] \ge 1- \frac{2}{z}.$$
\end{prevproof}

\end{appendix}
}
\end{document}